\documentclass[dvipsnames]{aa}  

\usepackage{graphicx}
\usepackage{txfonts}
\usepackage{url}
\usepackage{hyperref}
\hypersetup{
    colorlinks=true,
    linkcolor=blue,
    urlcolor=magenta,
    citecolor=blue
}
\usepackage{listings}
\usepackage{dsfont}
\usepackage{relsize}

\usepackage{orcidlink}

\begin{document}

   \title{ Analytic solution of Chemical Evolution Models with Type Ia SNe}

\subtitle{I. The disc bimodality in the [$\alpha$/Fe] vs. [Fe/H] plane and other applications}

   \author{P. A. Palicio
          \inst{1}\orcidlink{0000-0002-7432-8709}  \and E. Spitoni \inst{1,2}\orcidlink{0000-0001-9715-5727} \and A. Recio-Blanco\inst{1}\orcidlink{0000-0002-6550-7377} \and F. Matteucci\inst{2,3,4}\orcidlink{0000-0001-7067-2302} \and S. Peirani\inst{1,5}
        \and L. Greggio\inst{6} }

   \institute{Université Côte d’Azur, Observatoire de la Côte d’Azur, CNRS, Laboratoire Lagrange, France \and I.N.A.F. Osservatorio Astronomico di Trieste, via G.B. Tiepolo 11, I-34131, Trieste, Italy \and Dipartimento di Fisica, Sezione di Astronomia, Università di Trieste, via G.B. Tiepolo 11, I-34131, Trieste, Italy \and I.N.F.N. Sezione di Trieste, via Valerio 2, 34134 Trieste, Italy \and Sorbonne Université, CNRS, UMR7095, Institute d' Astrophysique de Paris, 98 bis Boulevard Arago, 75014 Paris, France \and I.N.A.F, Osservatorio Astronomico di Padova, Vicolo dell'Osservatorio 5, I-35122 Padova, Italy\\
              \email{pedro.alonso-palicio@oca.eu}
             }
             
    \date{Received March 2023; accepted July 2023}

 
  \abstract
    {In the last years, a significant number of works have focused on finding analytic solutions for the chemical enrichment models of galactic systems, including the Milky Way. Some of these solutions, however, cannot account for the enrichment produced by Type Ia SNe due to the presence of the delay time distributions (DTDs) in the models.}
   {We present a new analytic solution for the chemical evolution model of the Galaxy. This solution can be used with different prescriptions of the DTD, including the single and double degenerate scenarios, and allows the inclusion of an arbitrary number of pristine gas infalls.}
   {We integrate the chemical evolution model by extending the instantaneous recycling approximation with the contribution of Type Ia SNe. This implies an extra term in the modelling that depends on the DTD. For those DTDs that lead to non-analytic integrals, we describe them as a superposition of Gaussian, exponential and $1/t$ functions using a restricted least-squares fitting method. }
   {We obtain the exact solution for a chemical model with Type Ia SNe widely used in previous works, avoiding numerical integration errors. This solution can reproduce the expected chemical evolution of the $\alpha$ and iron-peak elements in less computing time than numerical integration methods. We compare the pattern in the [Si/Fe] vs. [Fe/H] plane observed by APOGEE DR17 with that predicted by the model. We find the low $\alpha$ sequence can be explained by a delayed gas infall. We exploit the applicability of our solution by modelling the chemical evolution of a simulated Milky Way-like galaxy from its star formation history. The implementation of our solution is released as a \textsc{python} package.}
   {Our solution constitutes a promising tool for the Galactic Archaeology and is able to model the observed trends in $\alpha$ element abundances versus $\rm [Fe/H]$ in the solar neighbourhood. We infer the chemical information of a simulated galaxy modelled without Chemistry.}

\titlerunning{Analytic solution of the CEM with Type Ia SNe}

\authorrunning{Palicio et al.}
   \maketitle
%
\section{Introduction}
\par The chemical evolution of galaxies describes the changes in the composition of the interstellar medium (ISM) produced by subsequent generations of stars. In this context, the analytic models are a powerful tool used to predict, among other examples, the evolution of the metallicity and the production of chemical elements on short time-scales in different galactic systems. 
\par Generally, analytic solutions for the chemical evolution of galaxies have been presented for the so-called 'Simple Model' \citep{tinsley1980}, which assumes: i) an initial mass function (IMF) not dependent on time, ii) the gas is well mixed at any time of the galaxy evolution (instantaneous mixing approximation, IMA) and iii) the lifetime of stars with mass $m \ge1\,\mathrm{M}_{\sun}$ is negligible compared to the timescale of the stars with mass $m<1\,\mathrm{M}_{\sun}$, whose longer lifetimes motivates the approximation of no contribution to the chemical enrichment of the ISM (see also the discussion in \citealt{matteucci2012,matteucci2021}). This assumption constitutes the instantaneous recycling approximation (IRA). In this framework, several analytic solutions for the evolution of the gas phase metallicity have been presented adding more complexity to the system: i.e.  infall of gas, galactic winds, radial gas flows, interactions between galaxies and galactic fountains \citep{matteucci1983, lacey1985,clayton1988, edmunds1990, recchi2008,spitoni2010,spitoni2015U,lilly2013,peng2015,kudritzki2015,kudritzki2021}. Under the IRA approximation, \citet{spitoni2017} presented the first analytic solution for the time evolution of the metallicity, gas mass fraction and total mass assuming an exponential infall rate of gas, although no predictions are made for iron abundance. \citet{beverage2021,spitoni_MZ2020,spitoni2021MDF} use this analytic solution to model the properties of the star forming and quenched galaxies. One of the major limitations of the models described above is the impossibility of obtaining realistic predictions for those elements produced on longer time-scales, such as iron.
%
%
\par Supernovae of Type Ia are considered the major producers of $^{56}$Fe, although a smaller fraction of this element is produced by core-collapse SNe. This was first demonstrated by \citet{Greggio1983} and then by \citet{matteucci1986}, who showed that with a normal IMF suitable for the Milky Way, the Type Ia SNe contribute by 70\% of the $^{56}$Fe enrichment in the solar vicinity. Therefore, understanding the Type Ia SN progenitors is key for the modelling of the iron production.
\par Historically, two main channels have been proposed for the formation of the Type Ia SNe: I) the single degenerate \citep[SD]{Whelan1973} and ii) the double degenerate \citep[DD]{Iben1994} scenarios. In the first channel the primary, intermediate mass star of a binary system evolves to produce a carbon oxygen white dwarf (WD) with a close companion. When the secondary star evolves, it fills its Roche Lobe promoting accretion on to the WD, which grows in mass and can explode reaching the Chandrasekhar limit (roughly 1.4~$\text{M}_{\odot}$). In the DD channel, when the secondary star fills its Roche Lobe the companion WD does not accrete the incoming material, forming instead a common envelope engulfing the two stars. This common envelope is eventually lost from the system, leaving behind a close double degenerate system. This system looses orbital energy by emitting gravitational waves \citep{LorenAguilar05}, leading to the final merging of the two WDs. If the total mass of the system exceeds the Chandrasekhar limit explosion can occur. Both channels imply a time delay between the formation of the progenitors and the final explosion that can be much larger that lifetime of massive stars. Thus, the instantaneous recycling approximation is no longer accurate for those elements produced mainly in Type Ia SNe, such as iron, requiring a more advance approach that accounts for the distribution of the mentioned time delay.
\par In the last years, more channels involving sub-Chandrasekhar mass have been proposed for Type Ia SNe to explain some peculiar cases \citep{Nomoto1982, Iben1991, Pakmor2012}. These alternatives, however, produce negligible effects on the $^{56}$Fe production (see the extensive discussion in \citealt{Palla2021} and references therein).
\par By relaxing the IRA approximation \citep{Chiosi1982}, the detailed chemical evolution model originally proposed by \citet{chiappini1997} can trace the dichotomy in the [$\alpha$/Fe] versus [Fe/H] diagram --- the so-called low- and high- $\alpha$ sequences --- observed in the Galactic disc \citep{Lee11, Haywood13, Haywood15, RecioBlanco:2014dd, Anders14, Nidever14, Hayden14, hayden2015, Bovy16, recioDR32022b} and subsequently explained by \citet{Kobayashi1998, Kobayashi2006, Fenner2002, noguchi2018, spitoni2019, spitoni2022, lian2020}. These works assume that the Galaxy has been formed by one or more separated accretion episodes, modelled by decaying time exponential-like infalls of gas.
\par \citet{vincenzo2017} provide a numeric solution for chemical evolution of the Galactic disc by assuming the IRA for chemical elements produced by massive stars and the delay time distribution formalism for the iron. They show the results for the solar neighbourhood for the one-infall model for the single degenerate (SD) scenario proposed by \citet[][]{matteucci2001}, as well as that considering the bimodal DTD of \citet[][]{mannucci2006}, in which one half of the Type Ia SNe are produced in prompt explosions ($\sim$ few Myr) while the remaining shows a wider time distribution ($\sim$ hundreds of Myr). \citet{vincenzo2017} find a good agreement between their predictions and the \citet{bensby2014} data for [O/Fe] and [Si/Fe] versus [Fe/H] assuming the DTD of \citet{matteucci2001}. 
\par \citet{weinbergE2017} avoid the numerical integration of the models with DTD by presenting the analytic solution for the evolution of the iron produced in Type Ia SN events. They are able to model the evolution of [Fe/H] and [$\alpha$/Fe] for three different star formation histories (constant, exponentially declining, linear-exponential). Their analysis, however, is restricted to a specific prescription for the delay time distribution of Type Ia SNe, which decreases exponentially with time. Similarly, \citet{pantoni2019,lapi2020} find the analytic solutions for the evolution of iron considering an exponential DTD. 
\par In this work, we present a new analytic solution for the chemical evolution of galactic systems accounting for the enrichment from Type Ia SNe. Compared to the numerical approach, the analytic solutions have the advantage of providing the exact abundance values at any evolutionary time, with no approximation errors and in a more direct fashion than the recurrent iteration over previous time steps. We consider a prescription for the DTD that extends these used in previous works \citep{weinbergE2017, pantoni2019, lapi2020}. We test our solution with the DTDs proposed by \citet{matteucci2001}, \citet[][]{greggio2005}, \citet{mannucci2006}, \citet[][]{totani2008}, \citet[][]{Pritchet08} and \citet[][]{strolger2004,strolger2005}. The paper is organised as follows: in Section \ref{sect_pres}, we present the chemical evolution model, the prescription for the IRA approximation and the adopted formalism for the Type Ia SNe enrichment (detailed in Appendix \ref{app_DTD_technical_section}). In Section \ref{sec_results} and Appendix \ref{app_sol}, we present the analytic solutions for different DTDs and apply them to the one and two infall scenarios. In Section \ref{sect_bimo}, we prove that the new solution can be a handful tool for the Galactic Archaeology. Using our analytic solution, we study the chemical dichotomy of the disc and compare it with that observed by APOGEE DR17 \citep{apogeedr172022}. We include similar prescriptions of the detailed two-infall models proposed by \citet{spitoni2019,spitoni2021} which were designed to reproduce APOKASC \citep{victor2018} and APOGEE DR16 \citep{Ahumada2019} data, respectively.
In Section \ref{Sec_Cosmo},  we model the iron and silicon abundances of a Milky Way-like galaxy from its SFR. The conclusions and future work are summarised in Section \ref{concl}.
\section{Model prescriptions}
\label{sect_pres}
In this Section, we present the main  model prescriptions. For a more complete discussion of Galactic chemical evolution assumptions and ingredients, we refer the reader to the review of \citet{matteucci2021} and the book of \citet{matteucci2012}.
\subsection{Useful quantities with IRA}\label{IRA_sec}
\par Under the assumption of the IRA and the instantaneous mixing approximation, the returned mass fraction $R$ indicates the total amount of mass restored to the ISM by a single stellar generation after $\sim10$~Gyr \citep{tinsley1980}. Given an IMF $\phi(m)$, $R$ can be computed as:
\begin{equation}
R = \frac{\int_{1 \, \text{M}_{\odot}}^{100\, \text{M}_{\odot}} (m - M_R) \phi (m) dm}{\int_{0.1 \, \text{M}_{\odot}}^{100\, \text{M}_{\odot}} m\phi (m) dm},
\label{eq:r}
\end{equation}
where $M_{R}$ is the mass of the stellar remnant. Similarly, the yield per stellar generation for the $X$ element $\langle y_X \rangle$ is defined as
\begin{equation}
\langle y_X \rangle = {1 \over {1 - R}} \int_{1 \, \text{M}_{\odot}}^{100\, \text{M}_{\odot}} m\,p_{X}(m)\,\phi (m) dm,
\label{eq:yield}
\end{equation}
where $p_{X}(m)$ is the ratio between the ejected mass of the element $i$ and newly produced by a star of mass $m$. As Eq. \ref{eq:yield} shows, only those stars with masses larger than 1 M$_{\odot}$ contribute to the chemical enrichment of the ISM, while the (1-$R$) term in the denominator accounts for the amount of mass locked up in stars of lower mass. Thus, $\langle y_X \rangle$ can be understood as the ratio between the ejected mass and remnant mass for the $X$ element in a single stellar generation. Both for $R$ and $\langle y_X \rangle$, the choice of the lower mass limit in their definitions does not significantly change their values \citep{tinsley1980}.
\subsection{Type Ia SNe and the  DTD formalism}
\label{DTD_sec}
\citet{greggio2005} proposed  a new formalism  for the Type SN Ia rate  based on the concept of the delay time distribution, namely the functional form which indicates how the SN Ia progenitors die as a function of time considering a  instantaneous starburst, i.e. a single stellar population. Given a SFR $\psi(t)$ and a delay time distribution $\rm DTD(t)$, the SN Ia rate at time $t$ is obtained as the following integral:
\begin{equation} 
\mathcal{R}_{\text{Ia}}(t) = C_{\text{Ia}} \int \limits_{\tau_{1}}^{\min(t,\tau_2)}   \text{DTD}_{\text{Ia}}(\tau) \, \psi(t-\tau) \, d\tau,  
\label{eq:Ia_DTD}
\end{equation}
where  $\tau_1$($\tau_2$) is the  minimum (maximum) time for the explosion of a Type Ia SN, and the normalisation constant $C_{\text{Ia}}$ is set to reproduce the observed present time Type Ia SN rate.
%
%
%
%
\begin{figure}
\begin{centering}
\includegraphics[width=0.45\textwidth]{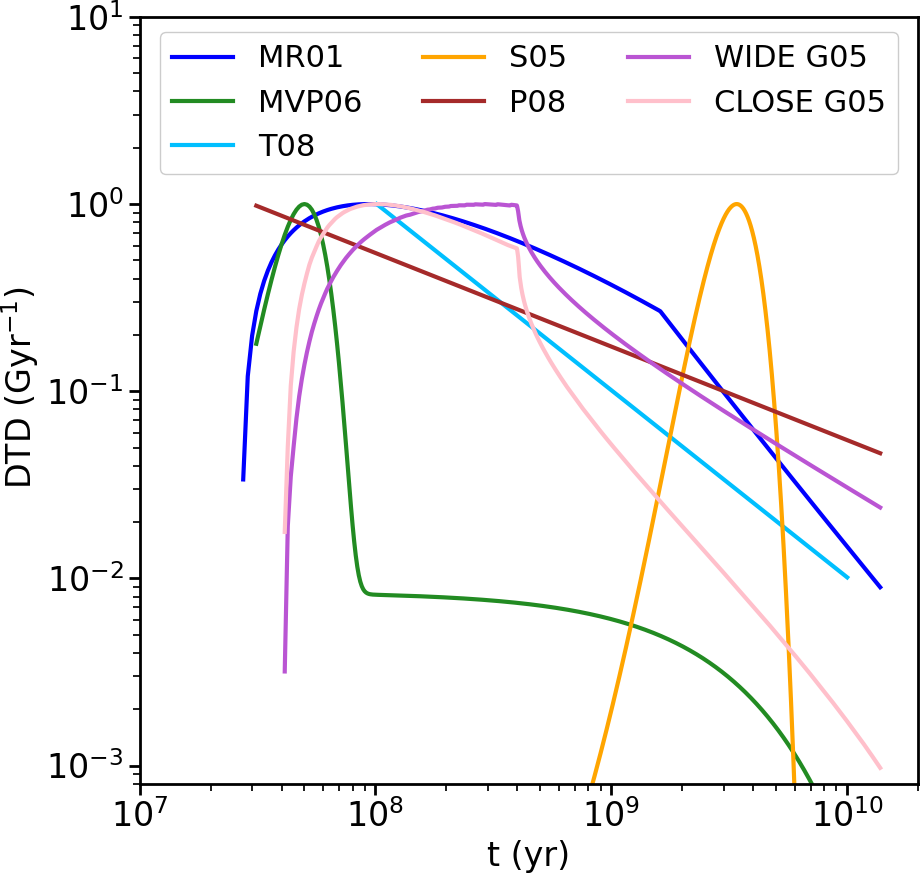}
\caption{Delay time distribution functions normalised to their own maximum value adopted in this work: MR01 \citep[][blue line]{matteucci2001}, WIDE and CLOSE G05 \citep[][purple and pink lines, respectively]{greggio2005},  MVP06  \citep[][green line]{mannucci2006}, S05 \citep[][yellow line]{strolger2005} and T08  \citep[][cyan line]{totani2008}.}
\label{Fig_DTDs}
\end{centering}
\end{figure}
\begin{figure}
\begin{centering}
\includegraphics[width=0.45\textwidth]{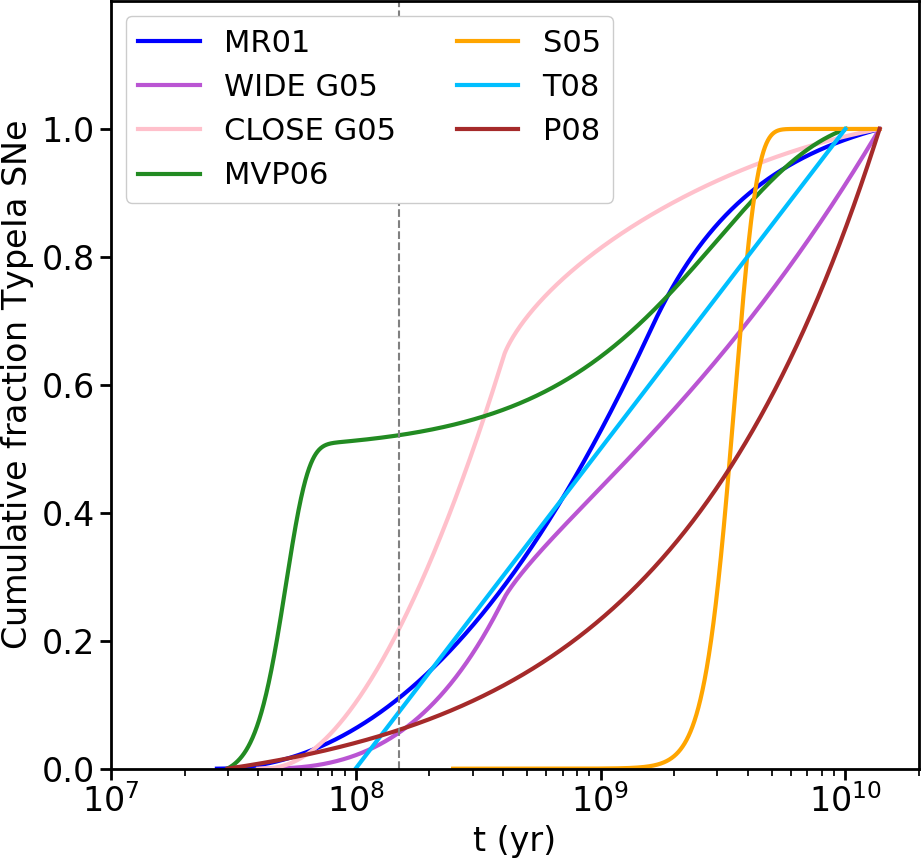}
\caption{Cumulative fraction of Type Ia SNe as a function of time for the DTDs used in this work: MR01 \citep[][blue line]{matteucci2001}, WIDE and CLOSE G05 \citep[][purple and pink lines, respectively]{greggio2005},  MVP06  \citep[][green line]{mannucci2006}, S05 \citep[][yellow line]{strolger2005} and T08  \citep[][cyan line]{totani2008}. Dashed vertical line denotes the reference time limit for the prompt of 150~Myr.}
\label{Fig_iDTDs}
\end{centering}
\end{figure}
%
%
%
%
%
\par In this work, we provide an analytic solution of the chemical evolution model considering different realisations of the DTD (Figs. \ref{Fig_DTDs} and \ref{Fig_iDTDs}): the one computed for the single degenerate scenario by \citet[][hereafter MR01]{matteucci2001}, those proposed by \citet[][hereafter G05]{greggio2005} for the WIDE and CLOSE double degenerate scenario, the DTDs derived empirically by \citet[][hereafter MVP06]{mannucci2006}, \citet[][hereafter T08]{totani2008} and \citet[][hereafter P08]{Pritchet08}, as well as the Gaussian DTD proposed by \citet[][herafter S05]{strolger2005} from the observed cosmic Type Ia SN rate. The terminology for the G05 DTDs is related to the distribution of the separation of the DD systems, which can be more or less populated at the low values, as resulting from respectively a less or more efficient transfer of orbital energy to the potential energy of the envelope. Correspondingly, the distribution of gravitational delays turns out more skewed towards the short delays in the CLOSE DD scheme, leading to steeper DTDs. We refer to \citet{greggio2005} for a detailed technical description of the two cases.
\par We consider the ample variety of formulations mentioned above because they all are based on astrophysical arguments. Notice that different DTDs can account for the observed Type Ia SN rates in external galaxies, within the current uncertainties \citep{Botticella17,greggio19}. Thus, the analytic formulations presented in this work for these DTDs can be used to construct models for the evolution of the iron abundance --or other element produced mainly in Type Ia SNe-- in other galaxies besides the Milky Way.
\par The DTD of MVP06 is described as a combination of a Gaussian and an exponential distribution, leading to the bimodality that characterises this DTD. On the contrary, S05 suggest a single Gaussian distribution with no prompt Type Ia SNe that peaks at 3–4 Gyr. The T08 and P08 DTDs are described by power law relations of the form $t^{-1}$ and $t^{-1/2}$, respectively, in which the T08 traces the slope of the WIDE G05 DTD. For a detailed explanation of the functional forms of the MR01 and G05 DTDs, we refer to \citet{matteucci2006}, \citet{greggio2005} as well as Sections 2.1 and 2.2 of \citet{bonaparte2013}.
\par Figure \ref{Fig_iDTDs} shows the cumulative fraction of Type Ia SNe from a single stellar population as a function of time. We consider the value of 150~Myr for the definition of the upper limit of the prompt regime (approximately an intermediate value among the time intervals given by \citealt{Acharova22, Aubourg08, Maoz10}). As we can note, there are significant differences among the prompt fractions of the DTDs: in the SD scenario the fraction of prompt Type Ia SNe is $\sim$~10\% \citep{bonaparte2013} while for the WIDE and CLOSE DD G05 scenarios these prompt fractions are $\sim$~6\% and $\sim$~22\%, respectively. For the MVP06 DTD this prompt fraction is substantially higher (50\%). The empirical T08 and P08 DTDs have prompt fractions similar to that of the WIDE DD case (approximately 6\% and 9\%, respectively).
%
\par In order to integrate analytically the chemical evolution model for the complex DTDs, we approximate them by a combination of truncated Gaussian, exponential and inverse of time ($\sim (t-\tau_0)^{-1}$) functions. Hence, the general expression for the fit of a DTD is
\begin{equation} 
\begin{split}
\text{DTD}(t) &=\sum_{i=1}^{N_G} A_{G,i}\exp\left({-\frac{(t - \tau'_i)^{2}}{2\sigma'^{2}_i}}\right)\cdot \mathds{1}_{\left[\tau_{1,G,i},\ \tau_{2,G,i}\right)}(t)
   \ +\ \\
&+ \sum_{i=1}^{N_E} A_{E,i}\exp{\left( \frac{-t}{\tau_{\text{D},i}}\right)}\cdot \mathds{1}_{\left[\tau_{1,E,i},\ \tau_{2,E,i}\right)}(t)
\ +\ \\
&+\sum_{i=1}^{N_I} A_{I,i}\frac{\tau_I}{t-\tau_0}\cdot \mathds{1}_{\left[\tau_{1,I,i},\ \tau_{2,I,i}\right)}(t),
\end{split}
\label{Eq_DTD_component}
\end{equation}
where we introduce $\tau_I=1.00$~Gyr just to keep the same units in all the amplitudes $A_{x,i}$. $N_G$, $N_E$ and $N_I$ are the number of Gaussian, exponential and $(t-\tau_0)^{-1}$ functions that characterise the DTDs, respectively\footnote{We use the mnemonic subindex naming ``G" for Gaussian, ``E'' for exponential and ``I'' for inverse.}. The value of the indicator function $\mathds{1}$ is one if the argument is within the interval of the sub-index and zero otherwise.
\par Note that within the formalism of Eq. \ref{Eq_DTD_component} the S05 and T08 DTDs emerge by setting $A_{E,i}$, $A_{I,i}=0$ and $A_{G,i}$, $A_{E,i}=0$, respectively. Similarly, a proper combination of $A_{G,i}$, $A_{E,i}$ with $A_{I,i}=0$ results in the MVP06 DTD. For those DTDs with a more complex functional form, like these of MR01, G05 and P05, the values of the amplitudes $A_{G,i}$, $A_{E,i}$, $A_{I,i}$ are determined by a restricted least-squares fitting that minimises the difference between the original DTD and Eq. \ref{Eq_DTD_component} (see Table \ref{tab_dtd}). Since the implementation of this method is rather technical, we refer to Appendix \ref{app_DTD_technical_section} for a more detailed explanation of this procedure.
%
%
%
%
%
\begin{table*}
\begin{center}
\tiny
\caption{Values of the parameters of the DTDs considered in this work. Without loosing generality, we can set $\tau_{\rm I}=1.00$ Gyr. For the MR01 DTD we assume $\gamma=0.5$ \citep{bonaparte2013}, while the WIDE and CLOSE G05 DTDs are defined by the tuple of parameters ($\tau_{n,x}=0.4$~Gyr, $\beta_a=0$) and ($\tau_{n,x}=0.4$~Gyr, $\beta_g$=-0.975), respectively \citep{greggio2005}. An online version of this table with higher decimal precision will be available. }
\label{tab_dtd}
\begin{tabular}{|c|ccccc|cccc|cccc|}
\hline
  \hline
& \multicolumn{13}{|c|}{DTD parameters}\\
 \hline

 & \multicolumn{5}{|c|}{Gaussian}&  \multicolumn{4}{|c|}{Exponential}& \multicolumn{4}{|c|}{Inverse}\\
DTD scenarios & $A_G$&$\tau'$&  $\sigma'$ & $\tau_{1, \, G}$& $\tau_{2, \,G}$ &$A_E$& $\tau_D$& $\tau_{1, \,E}$ &  $\tau_{2, \,E}$  & $A_I$& $\tau_0$ & $\tau_{1, \,I}$ &  $\tau_{2, \,I}$\\
 & [Gyr]$^{-1}$&[Gyr]&  [Gyr] & [Gyr]& [Gyr]&[Gyr]$^{-1}$& [Gyr]& [Gyr]&[Gyr]& [Gyr]$^{-1}$& [Gyr]&[Gyr]&[Gyr]\\
 \hline 
 & \multicolumn{5}{|c|}{}&  \multicolumn{4}{|c|}{}& \multicolumn{4}{|c|}{}\\
&-0.95&0.09&0.20&0.03&1.61&20.91&0.58&0.03&1.61&-0.06& 0.00& 0.03&1.61 \\ 
&0.19&0.23&0.13&0.03&1.61&-76.85&1.15&0.03&1.61&0.83&0.00&1.61&13.80 \\ 
&-0.06&1.28&0.20&0.03&1.61&61.41&1.73&0.03&1.61&/ &/ &/  &/ \\ 
 Single Degenerate &/ &/ &/ &/ &/ &65.76&2.30&0.03&1.61&/ &/ &/  &/ \\
 (\citealt[][]{matteucci2001},&/ &/ &/ &/ &/ &-67.93&2.88&0.03&1.61&/ &/ &/  &/ \\ 
MR01)&/ &/ &/ &/ &/ &0.03&1.79&1.61&13.80&/ &/ &/  &/ \\ 
&/ &/ &/ &/ &/ &-1.05&3.59&1.61&13.80&/ &/ &/  &/ \\ 
&/ &/ &/ &/ &/ &1.77&5.38&1.61&13.80&/ &/ &/  &/ \\ 
&/ &/ &/ &/ &/ &-1.13&7.18&1.61&13.80&/ &/ &/  &/ \\ 
   & \multicolumn{5}{|c|}{}&  \multicolumn{4}{|c|}{}& \multicolumn{4}{|c|}{}\\
\hline 
 & \multicolumn{5}{|c|}{}&  \multicolumn{4}{|c|}{}& \multicolumn{4}{|c|}{}\\
&-0.01&0.09&0.09&0.04&0.40&3.44&3.19&0.04&0.40&-0.06&0.00&0.04&0.40\\ 
&-0.06&0.40&0.10&0.40&13.80&-2.02&6.38&0.04&0.40&0.04&0.35&0.40&13.80\\ 
&0.08&0.40&0.20&0.40&13.80&0.27&2.80&0.40&13.80&/ &/ &/ &/ \\ 
WIDE Double Degenerate&0.02&0.40&0.30&0.40&13.80&-0.65&5.36&0.40&13.80&/ &/ &/ &/ \\ 
\citep[][WIDE G05]{greggio2005}&-0.09&0.40&0.40&0.40&13.80&0.64&7.91&0.40&13.80&/ &/ &/ &/ \\ 
&0.12&0.40&0.50&0.40&13.80&1.52&10.47&0.40&13.80&/ &/ &/ &/ \\ 
&/ &/ &/ &/ &/ &-3.31&13.02&0.40&13.80&/ &/ &/ &/ \\ 
&/ &/ &/ &/ &/ &1.69&15.57&0.40&13.80&/ &/ &/ &/ \\ 
& & & & & & & & & & & & & \\
\hline
& \multicolumn{5}{|c|}{}&  \multicolumn{4}{|c|}{}& \multicolumn{4}{|c|}{}\\
&-1.84E-2&0.10&0.02&0.04&0.40&9.92&0.27&0.04&0.40&-0.13&0.00&0.04&0.40\\ 
&0.07&0.24&0.02&0.04&0.40&-0.23&0.54&0.04&0.40&0.50&0.33&0.40&13.80\\ 
&0.08&0.30&0.02&0.04&0.40&-34.92&0.81&0.04&0.40&-1.18&0.30&0.40&13.80\\ 
&0.02&0.18&0.02&0.04&0.40&29.04&1.09&0.04&0.40&0.68&0.25&0.40&13.80\\ 
&0.04&0.36&0.02&0.04&0.40&0.22&0.14&0.40&13.80&/ &/ &/ &/ \\ 
CLOSE Double Degenerate&0.03&0.08&0.02&0.04&0.40&1.71&0.24&0.40&13.80&/ &/ &/ &/ \\ 
\citep[][CLOSE G05]{greggio2005}&0.14&0.40&0.10&0.40&13.80&-0.04&0.52&0.40&13.80&/ &/ &/ &/ \\ 
&2.27E-3&3.00&1.32&0.40&13.80&0.03&0.93&0.40&13.80&/ &/ &/ &/ \\ 
&-2.97E-2&1.32&1.81&0.40&13.80&0.10&2.17&0.40&13.80&/ &/ &/ &/ \\ 
&-1.68E-3&6.00&2.07&0.40&13.80&/ &/ &/ &/ &/ &/ &/ &/ \\ 
&-3.07E-3&4.60&1.47&0.40&13.80&/ &/ &/ &/ &/ &/ &/ &/ \\ 
&-9.54E-5&7.80&1.87&0.40&13.80&/ &/ &/ &/ &/ &/ &/ &/ \\
& \multicolumn{5}{|c|}{}&  \multicolumn{4}{|c|}{}& \multicolumn{4}{|c|}{}\\
\hline
  & \multicolumn{5}{|c|}{}&  \multicolumn{4}{|c|}{}& \multicolumn{4}{|c|}{}\\
Empirical bimodal distribution&19.95&0.05&0.01&0.03&10.05&0.17&3.00&0.03&10.05& /&/&/&/\\
  \citep[][MPV06]{mannucci2006}& &&  && && && &&&&\\
& &&  && && && &&&&\\
\hline 
 & \multicolumn{5}{|c|}{}&  \multicolumn{4}{|c|}{}& \multicolumn{4}{|c|}{}\\
Empirical  $\propto t^{-1}$& /&/&  / &/&/&/& / &/&/ &1.00&0.00&0.10&10.00\\
  \citep[][T08]{totani2008}& &&  && &&&&  &&&&\\
   & &&  && &&&&  &&&&\\
\hline 
& & & & & & & & & & & & & \\
&-0.15&3.5E-3&0.10&0.03&13.80&0.69&5.56&0.03&13.80&-0.03&0.03&0.03&13.80\\ 
Empirical $\propto t^{-1/2}$&/ &/ &/ &/ &/ &-3.38&11.07&0.03&13.80&0.31&0.02&0.03&13.80\\ 
\citep[][P08]{Pritchet08}&/ &/ &/ &/ &/ &5.57&16.58&0.03&13.80&-0.83&0.01&0.03&13.80\\ 
&/ &/ &/ &/ &/ &-2.75&22.09&0.03&13.80&0.60&0.01&0.03&13.80\\ 
& & & & & & & & & & & & & \\
\hline
 & \multicolumn{5}{|c|}{}&  \multicolumn{4}{|c|}{}& \multicolumn{4}{|c|}{}\\
Empirical Gaussian&1.00&3.40&0.68&0.25&13.80&/& / &/& /& /&/&/&/\\
  \citep[][S05]{strolger2004,strolger2005}& &&  && &&&&  &&&&\\
    & &&  && &&&&  &&&&\\
\hline
\end{tabular}
\end{center}
\end{table*}
\begin{table*}
\begin{center}
\tiny
\caption{Model parameters considered in
the proposed chemical evolution model distinguishing between
"Galaxy Model", "IRA", "Type Ia SNe \& DTD" quantities.   In the last rows the  parameters adopted in the solution reported in Appendix \ref{app_sol}  are also indicated. }
\label{tab_sel}
\begin{tabular}{|c|cc|c|}
\hline
  \hline
& \multicolumn{2}{|c|}{Model parameters}& Description\\
 & Name& Dimension&  \\
 \hline
 & \multicolumn{2}{|c|}{}& \\
 & $N$&1& Number of infall episodes\\
 & $\tau_j$& [Gyr]& Time-scale of gas accretion for the $j^{th}$ infall episode  \\
 & $\rm t_j$& [Gyr]& Starting time of the $j^{th}$ infall episode  \\
 & $A_j$& [M$_{\odot}$ pc$^{-2}$ Gyr$^{-1}$]& Normalisation coefficient of the $j^{th}$ infall episode \\
Galaxy & $\sigma_{Aj}$& [M$_{\odot}$ pc$^{-2}$]& Total accreted  surface mass density for the $j^{th}$ infall    \\
Model& $\omega$& 1& Wind loading factor   \\
& $\nu_L$& [Gyr$^{-1}$]& Star-formation efficiency  \\
& $\sigma_{gas}$& [M$_{\odot}$ pc$^{-2}$]& Total Surface gas density  \\
& $\sigma_{\star}$& [M$_{\odot}$ pc$^{-2}$]& Total Surface stellar mass density  \\
& $\sigma_X(0)$ &[M$_{\odot}$ pc$^{-2}$] & Initial surface mass density of the element $X$\\

 & \multicolumn{2}{|c|}{}& \\
\hline   
 & \multicolumn{2}{|c|}{}& \\
& $R$& 1& Recycling fraction \\
IRA& $ \left< y_{X} \right> $& 1& Yield per stellar generation for the element X \\
 & \multicolumn{2}{|c|}{}& \\

\hline 
 & \multicolumn{2}{|c|}{}& \\
& $\left< m_{X,\text{Ia}}  \right>$ &[M$_{\odot}$] & Average amount of X synthesized by each single Type Ia SN event. \\
 &  $C_{Ia}$ &[M$_{\odot}^{-1}$]& Normalisation constant for the Type Ia SNe rate.\\
&  $A_G$ &[Gyr$^{-1}$] & Amplitude of the Gaussian term in the DTD.\\

 Type Ia SNe&  $A_E$ &[Gyr$^{-1}$] & Amplitude of the exponential term in the DTD.\\
\&&  $A_I$ &[Gyr$^{-1}$] & Amplitude of the $t^{-1}$ term in the DTD.\\
DTD Model &  $\sigma'$ &[Gyr] &  Width of the  Gaussian DTD.\\
 &  $\tau'$ &[Gyr] & Median of the Gaussian DTD.\\
 &  $\tau_D$ &[Gyr] & Timescale of the exponential DTD.\\
 &  $\tau_0$ &[Gyr] & Offset of the inverse DTD. It must satisfy $\tau_0<\tau_{I,1}$.\\
 &  $\tau_I$ &[Gyr] & Characteristic time of the inverse DTD (set to 1~Gyr).\\
 & \multicolumn{2}{|c|}{}& \\
\hline 
 & \multicolumn{2}{|c|}{}& \\
 &  $\Delta t_j$ &[Gyr] & $t-t_j$\\
 &  $\alpha$ &[Gyr$^{-1}$] & $\nu_L(1+\omega-R)$ \\ 
Solution & $\beta_j$ &[Gyr$^{-1}$] & $\alpha -1/\tau_j$\\
parameters &  $\eta_j$ &[Gyr] & $\tau'+\sigma'^2 /\tau_j$\\
 &  $\eta_\alpha$ &[Gyr] & $\tau'+\sigma'^2 \alpha$\\
 & $\sigma_{tot}$& [M$_{\odot}$ pc$^{-2}$]& $\sigma_g+\sigma_{\star}$ \\
  & \multicolumn{2}{|c|}{}& \\
\hline 
\end{tabular}
\end{center}
\end{table*}
%
%
%
%
%
%
%
%
%
\subsection{Chemical evolution equations}
\label{sec_che_eq}
We consider a one zone chemical evolution model, assuming the following form for the Kennicutt-Schmidt law \citep{schmidt1959, Kennicutt89} law for the SFR:
\begin{equation}
\label{Eq_psi_gas}
\psi(t)=\nu_L  \, \sigma_{gas}(t), 
\end{equation} 
where $\nu_L$ is the star formation efficiency (SFE) and has the dimension of [Gyr$^{-1}$]. As in \citet{spitoni2017} and \citet{vincenzo2017}, we consider galactic winds proportional to the SFR: 
\begin{equation}
W(t)=\omega\,\psi(t).
\end{equation}
\par In the scenario proposed by several works in literature \citep[e.g.,][]{Chiosi1980, Boissier2000, SchBinney09, Andrews2017}, the galaxy has been formed out by the accretion of distinctive  exponential infall events. Here, we provide analytic solutions for the chemical evolution of a system built up by $N$ infalls, in which the total gas accretion rate can be expressed as
\begin{equation}\label{infall_eq} 
I(t)= \sum_{j=1}^{N}  A_j\exp{\left(-\dfrac{\Delta t_j}{\tau_j}\right)}\,\theta(\Delta t_j),
\end{equation} 
where the j-th infall starts at time $t_j$ and is characterised by the timescale $\tau_j$; while the amplitude $A_j$ tunes the amount of gas accreted due to the j-th infall. In order to simplify the notation, we denote $t-t_j$ as $\Delta t_j$ and the Heaviside step function as $\theta$.
\citet{vincenzo2017} 
provided the following analytic  expression for  the star formation history of a galactic system formed by  the accretion of $N$ separate infalls  characterised by  exponential rate decays in presence of the IRA and the  \citet{schmidt1959} law for the SFR:
\begin{flalign} \label{eq:sfr_sol}
 \psi(t) &=\nu_{\text{L}}\mathlarger{\mathlarger{\sum}}_{\substack{j=1\\ 
                  \tau_j \neq \alpha^{-1}}}^{N} { \dfrac{\,A_{j}\tau_{j}}{\alpha\tau_j - 1}\,\,\Big[ \exp{\left(-\Delta t_j/\tau_{j}\right)} - \exp{\left(-\alpha \Delta t_j\right)} \Big] \theta(\Delta t_j)   }  \nonumber \\
& +\nu_{\text{L}}\sigma_{\text{gas}}(0)\,\exp{\left(-\alpha t\right)}\,\theta(t)  \\
 &+\nu_{\text{L}} \mathlarger{\mathlarger{\sum}}_{\substack{j=1\\
                  \tau_j = \alpha^{-1}}}^{N}{ A_{j}\, \Delta t_j \cdot \, \exp{\left(-\alpha \Delta t_j\right)}\cdot \theta(\Delta t_j)},\nonumber
\end{flalign}
where $\alpha=\nu_L(1+\omega-R)$, as indicated in Table \ref{tab_sel}. Compared to eq. 13 in \citet{vincenzo2017}, we include the additional summation term in $\sim A_j \Delta t_j$ that corresponds to the particular case in which $\tau_j \rightarrow{\alpha}^{-1}$. Although this term is necessary to provide the full general solution, we can ignore it hereafter since we do not make use of any $\tau_j=\alpha^{-1}$ in this work. The contribution of this term, however, can be found in Appendix \ref{app_sol}. Finally, the equation for the evolution of the surface gas density for the $X$-element $\sigma_{X}(t)$ with the Type Ia SN contribution reads:
\begin{eqnarray} \label{eq_to_solve}
 \dfrac{d\sigma_{X}(t)}{dt} = \overbrace{ - \alpha\, \sigma_{X}(t)  + \left\langle y_{X} \right\rangle \big( 1-R \big)\, \psi(t)} ^{\text{{IRA terms}}} + \overbrace{\left\langle m_{X,\text{Ia}} \right\rangle \mathcal{R}_{\text{Ia}}(t)}^{\text{{Type Ia SNe term}}},
\end{eqnarray}
where  $\left\langle m_{X,\text{Ia}} \right\rangle$ is the mass of the element $X$ synthesized by each single Type Ia SN explosion. 
\par This is a first order inhomogeneous differential equation whose solution can be computed analytically if the SFR $\psi(t)$ has the form given by Eq. \ref{eq:sfr_sol}. This solution is summarised in Appendix \ref{app_sol}.
\section{Results}
\label{sec_results}
In Section \ref{sec_new},  we present the new analytic solutions for the temporal evolution of different chemical elements considering the iron produced by different DTD prescriptions. We also test the effects of different DTD prescriptions for the one-infall scenario on the [$\alpha$/Fe] versus [Fe/H] abundance ratios (Section \ref{sec_1infall}) and on the metallicity distribution functions (Section \ref{sec_MDF}). We refer the reader to  Appendix \ref{app_sol} for the detailed explanation of the analytic form of the solutions and to the {\tt ChEAP}\footnote{\url{https://bitbucket.org/pedroap/cheap/src/master/}} ({\tt Ch}emical {\tt E}volution {\tt A}nalytic {\tt P}ackage) repository for its implementation in the \textsc{python} language.
\subsection{The new analytic solution}
\label{sec_new}
\par In Appendix \ref{app_sol}, we present the analytic expression for the Type Ia SN rates $\mathcal{R}_{\text{Ia}}(t)$ and the surface mass density $\sigma_X$ of the element $X$, which can be written as the sum of the contribution of the IRA and Type Ia SNe enrichment as:
\begin{equation}
 \sigma_X(t)=  \sigma_{X, \,  \text{IRA}}(t)+\sigma_{X, \, \text{Ia}}(t).
 \label{eq:diff_equation2}
\end{equation} 
Similarly, $\sigma_{X, \, \text{Ia}}(t)$ can be separated into terms that depend on the Gaussian  $(\sigma_{X, Ia, G})$, exponential  $(\sigma_{X, Ia, E})$  and inverse time $(\sigma_{X, Ia, I})$ DTDs:
\begin{equation}
\sigma_{X, Ia}(t)= \sum^{N_G} \sigma_{X, Ia, G}(t)+ \sum^{N_E} \sigma_{X, Ia, E}(t)+\sum^{N_I} \sigma_{X, Ia, I}(t).
\label{eq_sumIA}
\end{equation}
\par Table \ref{tab_sel} summarises all the parameters considered in the proposed chemical evolution model, distinguishing between "Galaxy Model", "IRA", "Type Ia SNe \& DTD" quantities. Furthermore, we provide some useful definitions to simplify the analytic expressions.
\par Once $\sigma_X(t)$ is known, we compute the abundance ratio [$X$/Fe] as:
\begin{equation}
\label{Eq_X_Fe}
[X/\text{Fe}]= \log_{10} \left( \frac{\sigma_X}{\sigma_{\text{Fe}}}         \right)- \text{SV}^{X}_{\text{Fe}},
\end{equation}
where $\text{SV}^{X}_{\text{Fe}}$ is a scaling factor derived from the solar reference values of \citet{asplund2009}. For the particular case of the iron, its abundance is computed as
\begin{equation}
\label{Eq_Fe_H}
[\text{Fe}/\text{H}]= \log_{10}  \left( \frac{\sigma_{\text{Fe}}}{0.75 \cdot \sigma_{gas}}         \right) - \text{SV}^{\text{Fe}}_{\text{H}},
\end{equation}
in which $\sigma_{gas}$ is given by Eq. \ref{Eq_psi_gas} through Eq. \ref{eq:sfr_sol}. From the Big Bang nucleosynthesis we assume the hydrogen comprises the 75\% of the gas mass (factor 0.75 in Eq. \ref{Eq_Fe_H}). Also, we assume the H abundance in mass does not change significantly during the Galactic evolution.
\begin{figure}
\begin{centering}
\includegraphics[width=0.45\textwidth]{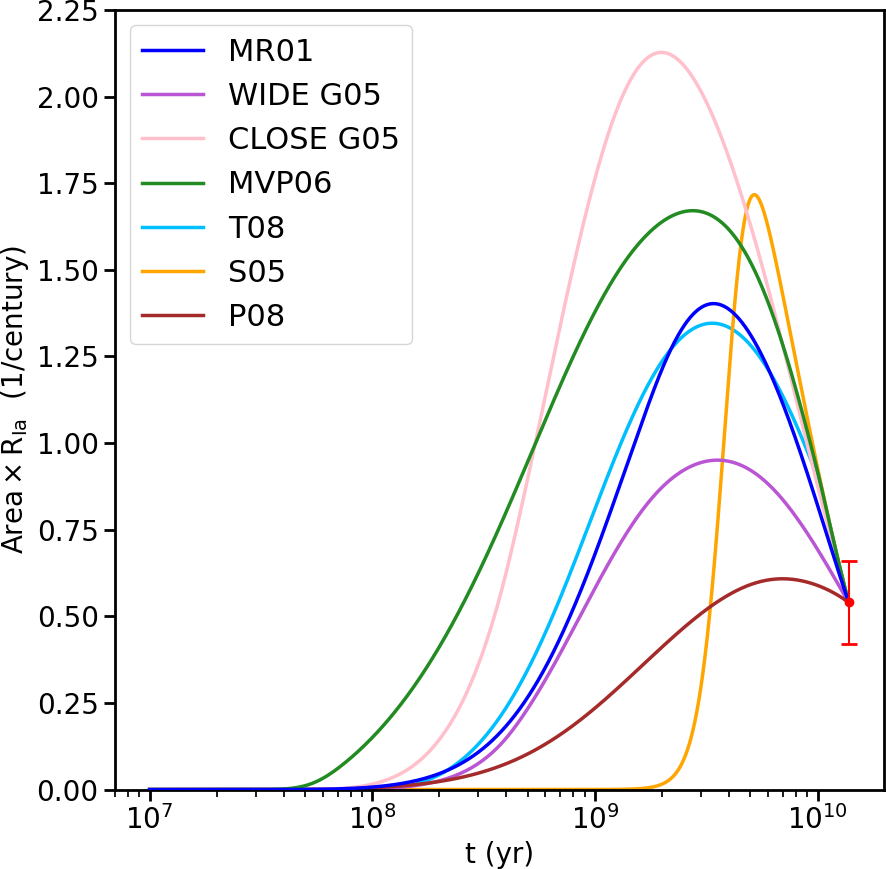}
\caption{Predicted Type Ia SN rates for the different DTDs considered in this work (see Fig. \ref{Fig_DTDs}) assuming the one-infall model introduced in Section \ref{sec_1infall}. The area in the vertical axis refers to that of the Galactic disc region ($3$~kpc$<R<20$~kpc). The rates are expressed in SN events per century and normalized (through the $C_{Ia}$) to reproduce the present time Type Ia SN rate in the Milky Way \citep[][filled red circle and error bar]{li2011}.}
\label{SNIrate_1}
\end{centering}
\end{figure}
\subsection{Testing the new analytic solutions: effects of the DTD on the one-infall model }
\label{sec_1infall}
In this Section we show the effects of different DTD prescriptions on the simplest case of the one-infall scenario at $t_1=0$~Gyr. This model has been widely used in the past to describe the thin disc of our Galaxy \citep{spitoni2015,spitoni2D2019, grisoni2017,grisoni2018}. As in \citet{vincenzo2017}, we study the evolution of oxygen, silicon and iron by assuming $\left<y_{\text{O}}\right>=1.022\times10^{-2}$, $\left<y_{\text{Si}}\right>=8.5\times10^{-4}$, 
$\left<y_{\text{Fe}}\right>=5.6\times10^{-4}$ and the returned fraction $R=0.285$. These values are derived from the \citet{kroupa1993} IMF and the collection of nucleosynthesis yields suggested by \citet[][see also \citealt{vincenzo2016}]{romano2010}. For the stellar yields of Type Ia SNe we make use of those from \citet{iwamoto1999}.
\par We consider the same set of parameters as in \citet{vincenzo2017}. This implies a total surface mass density in the solar neighbourhood of $\sigma_{\text{tot}}  = 54\,\text{M}_{\sun}\,\text{pc}^{-2}$ and also used widely in other works \citep[e.g.,][]{spitoni2015,spitoni2011}, a star formation efficiency $\nu_{\text{L}}=2\,\text{Gyr}^{-1}$, an infall time scale for the gas mass accretion of $\tau_1=7\,\text{Gyr}$ and a mass loading factor $\omega=0.4$. In contrast to \citet{vincenzo2017}, we set the value of $C_{Ia}$ (see Eq.  \ref{eq:Ia_DTD}) by imposing the predicted present-day SN Type Ia rate of $0.54 \pm 0.12$~events per century \citep{li2011} in the disc region (3~kpc~$<R<$~20~kpc). Thus, we assume the solar neighbourhood is representative of that annular region. No initial amount of gas is assumed for the Milky Way ($\sigma_{gas}=0~\rm~M_{\odot}\,pc^{-2}\,Gyr^{-1}$).
\par Figure \ref{SNIrate_1} shows the temporal evolution of Type Ia SN rates for the seven DTDs considered in this study. The rate $\mathcal{R}_{\text{Ia}}(t)$ computed with the S05 DTD peaks at later evolution time with respect to the majority of the other distributions because of the lack of prompt Type Ia SNe. However, the variation of $\mathcal{R}_{\text{Ia}}$  among the DTDs  does not depend only on the shape of the DTD but also on its ``convolution'' with the star formation history (Eq. \ref{eq:Ia_DTD}).
\par
\begin{figure*}
\begin{centering}
\includegraphics[width=0.85\textwidth]{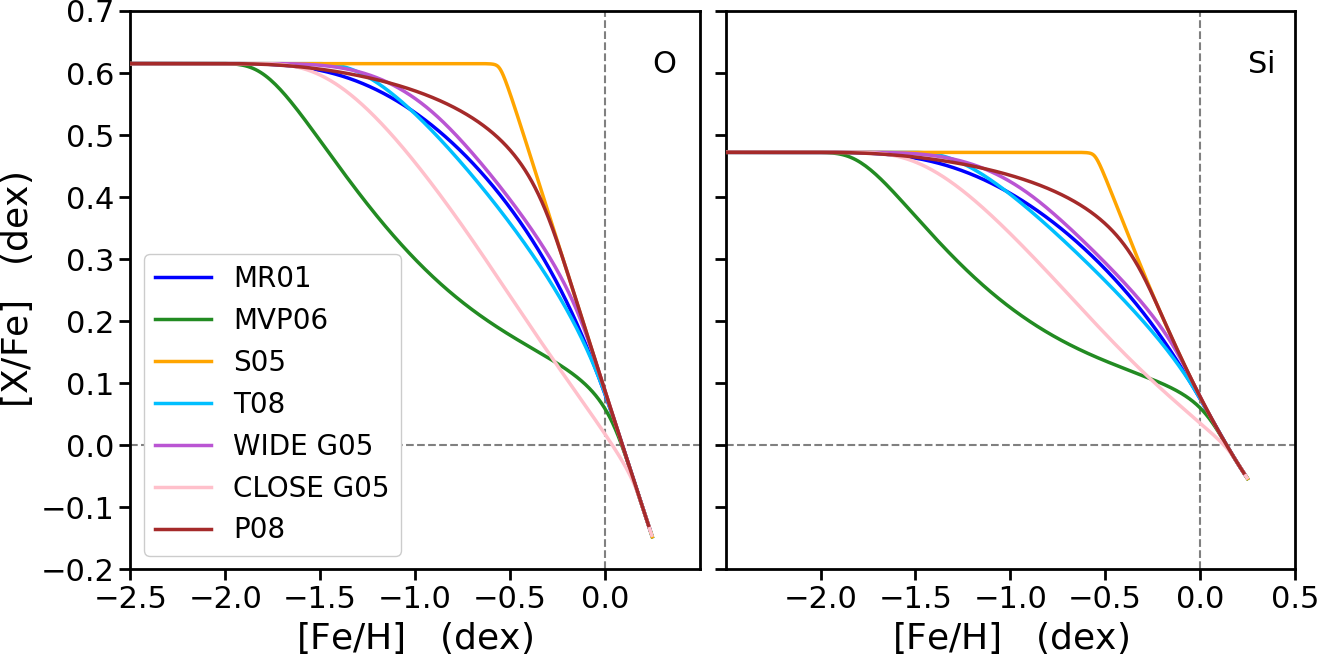}
\caption{ Evolution of [O/Fe] (left panel) and [Si/Fe] (right panel) versus [Fe/H] predicted by the one-infall model for seven different DTDs (solid curves). See Section \ref{sec_1infall} for the details of the model.}
\label{Fig_OSI_1infall}
\end{centering}
\end{figure*}

 In Fig. \ref{Fig_OSI_1infall} we show the predicted [O/Fe] (left panel) and [Si/Fe] (right panel) versus [Fe/H] for the seven DTD prescriptions. As already noted in \citet{matteucci2009}, the DTD of S05 shows the longest plateau in the [$\alpha$/Fe] versus [Fe/H] diagram because no prompt Type Ia SNe is present. The predicted abundance ratios agree with the \citet{matteucci2009} and \citet{vincenzo2017} results. We have checked that the fraction $\sigma_X/\sigma_{\text{Fe}}$ predicted by our solution --- for a generic chemical element $X$ --- equals the ratio of the yields $\langle y_X\rangle/\langle y_{\text{Fe}}\rangle$ in the limit $t\rightarrow{0}$~Gyr.
\begin{figure*}
\begin{centering}
\includegraphics[width=0.83\textwidth]{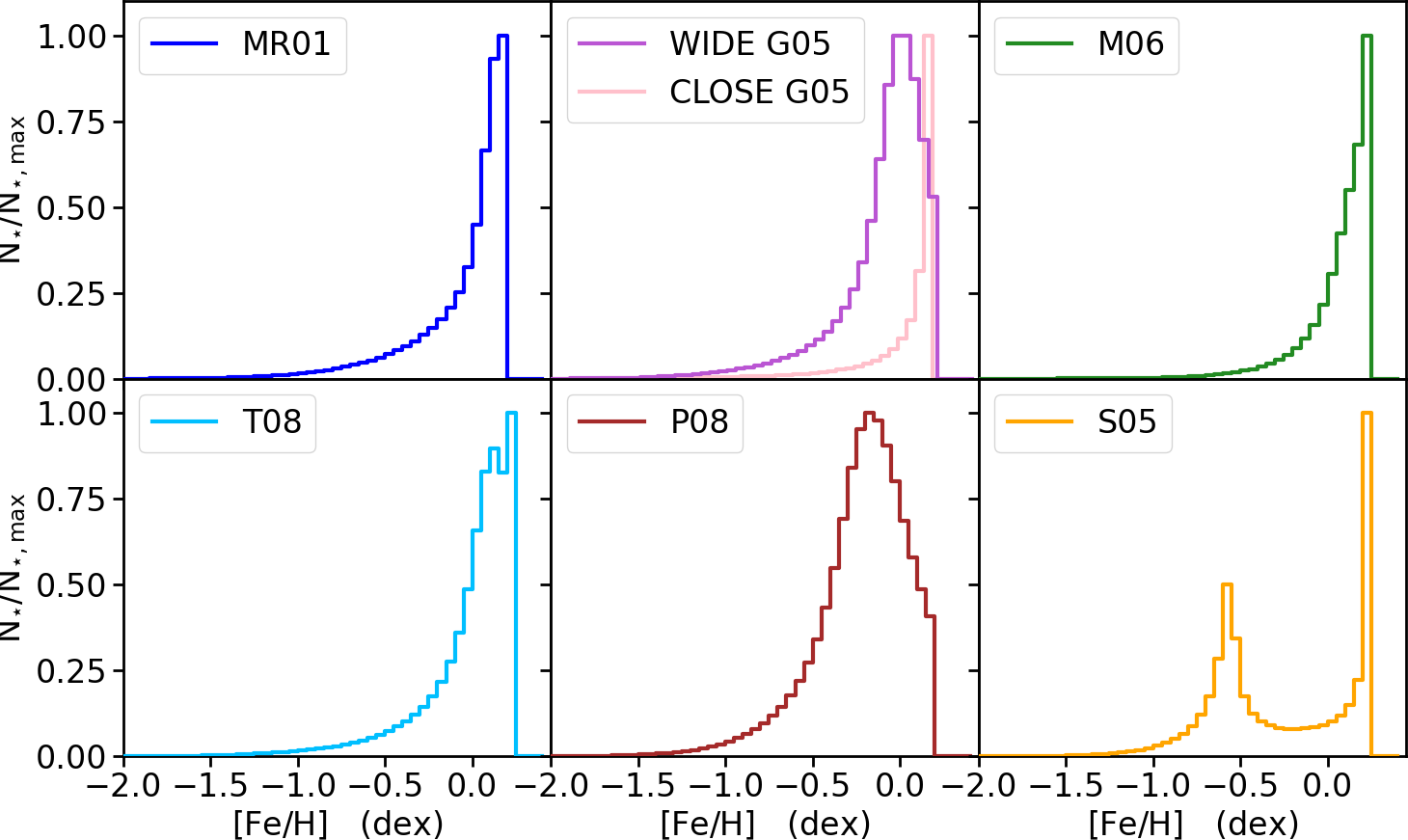}
\caption{ Normalised MDFs predicted by the one-infall model (see  Section  \ref{sec_1infall}) for different DTDs. Color code is the same as in Fig. \ref{Fig_OSI_1infall}. }
\label{MDF_1infall}
\end{centering}
\end{figure*}
\begin{figure}
\begin{centering}
\includegraphics[width=0.47\textwidth]{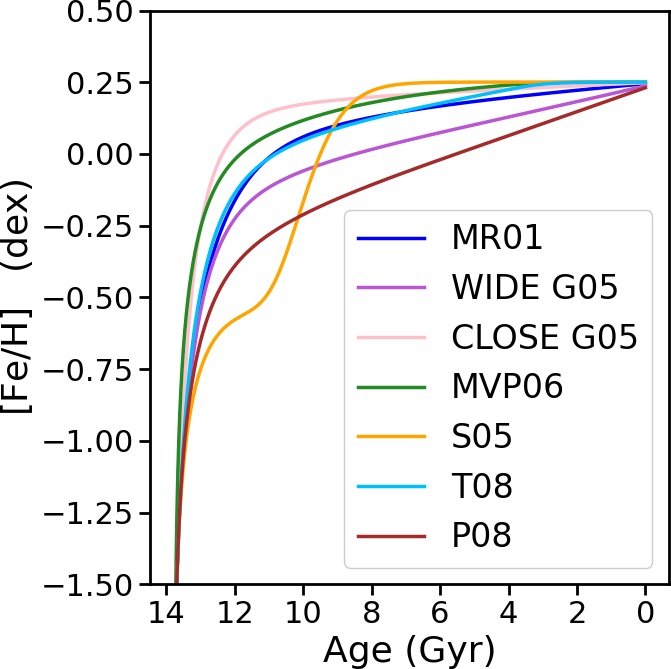}
\caption{  Age-metallicity relation for the one-infall model (see  Section  \ref{sec_1infall}) for different DTDs. Color code is the same as in Fig. \ref{Fig_OSI_1infall}. }
\label{Fig_AgeMet_1infall}
\end{centering}
\end{figure}
\begin{figure}
\begin{centering}
\includegraphics[width=0.47\textwidth]{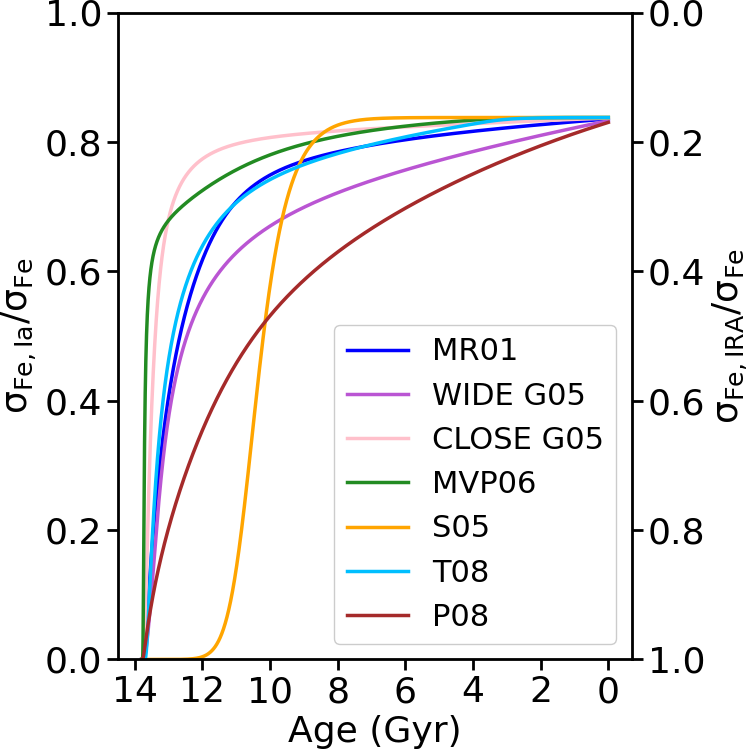}
\caption{  Fraction of iron produced by Type Ia SNe as a function of age (left vertical axis) assuming the MR01 (blue line), WIDE and CLOSE G05 (purple and pink, respectively), MVP06 (green), S05 (orange), T08 (cyan) and P08 DTDs (brown). The right vertical axis indicates the fraction of iron produced by Type II SNe. }
\label{Fig_Sigmat_1infall}
\end{centering}
\end{figure}
\subsection{The metallicity distribution function}
\label{sec_MDF}
\par The proposed analytic solution can be used to predict the metallicity distribution function\footnote{We consider the distribution of [Fe/H] as a proxy of the global metallicity distribution.} (MDF). Given the star formation rate $\psi(t)$, the amount of stellar mass $dM$ formed in an area $dA$ in the time interval $[t$,~$t+dt]$ is
\begin{equation}
    \label{Eq_dM}
    dM = \psi(t)\cdot dA\cdot dt.
\end{equation}
\par By definition of initial mass function $\phi(m)$, the total number of stars with masses $m_{\star, \rm min}\leq m < m_{\star}$ resulting from a single formation event is
\begin{equation}
    \label{Eq_dN}
    N_{\star} = k \int_{m_{\star, \rm min}}^{m_{\star}}\phi(m) dm,
\end{equation}
where the value of the constant $k$ is determined by the total mass of the population $M_{\star,\rm tot}$ as
\begin{equation}
    \label{Eq_kimf}
    k = \frac{M_{\star,\rm{ tot}}}{\int_{m_{\star, \rm min}}^{m_{\star, \rm max}}m\cdot\phi(m) dm}.
\end{equation}
\par Note that the integration limits in Eq. \ref{Eq_kimf} includes all the possible stellar masses, without excluding sources more massive than $m_{\star}$ as in Eq. \ref{Eq_dN}. Thus, for a population of mass $M_{\star,\rm {tot}}=dM$, the total number of stars within $m_{\star, \rm min}\leq m < m_{\star}$ per unit of area $dN_{\star}/dA$ is
\begin{equation}
    \label{Eq_dN_long}
    \frac{dN_{\star}}{dA} = \psi(t) \frac{\int_{m_{\star, \rm min}}^{m_{\star}}\phi(m) dm}{\int_{m_{\star, \rm min}}^{m_{\star, \rm max}}m\cdot\phi(m) dm} \cdot dt=f\cdot \psi(t) \cdot dt.
\end{equation}
\par Since the IRA approximation assumes all the stars more massive than the Sun die immediately, we must consider $m_{\star}=1 M_{\odot}$ in the numerator of Eq. \ref{Eq_dN_long}. For $m_{\star, \rm min}$ and $m_{\star, \rm max}$ we use the values 0.1  $\text{M}_{\odot}$ and 100  $\text{M}_{\odot}$, respectively, to be consistent with the $R$ and $\langle y_X\rangle$ values adopted in this study.
\par We can construct the MDF from Eq. \ref{Eq_dN_long} by integrating $dN_{\star}$ within the limits of each bin in metallicity. This procedure, however, generally requires solving transcendental equations to get the integration limits as a function of the metallicity. This is specially complicate when several infalls are included, since one may need to account for multiple branches of $t(\rm{[Fe/H]})$. A more practical approach is performed thanks to the following numerical integration
\begin{equation}
\label{Eq_num_MDF}
\begin{split}
    &MDF_i = \int_A \int_0^{t_G} dN_{\star} = \\
    &= f \cdot \int_{A} \int_0^{t_G} \psi(t)\cdot  \mathds{1}_{\left[[M/H]_i,\ [M/H]_{i+1}\right)}([M/H](t))\cdot dt\cdot dA,
\end{split}
\end{equation}
where the left hand side term is the number of stars in the $i-th$ metallicity bin $[M/H]_i\leq [M/H](t)<[M/H]_{i+1}$, $A$ is the integration area and $t_G=13.8$~Gyr. Since Eq. \ref{eq_to_solve} has no spatial dependence, we can substitute the integral in $dA$ by the total area $A$.
%
%
\par In Fig. \ref{MDF_1infall}, we show the MDFs predicted by the one-infall model for the different DTDs. We can see the MR01, CLOSE G05, MVP06 and T08 DTDs results in similar MDFs, while the S05 DTDs shows a peak at $\rm [Fe/H]\approx-0.6$~dex. The MDFs computed with the WIDE G05 and P08 DTDs peak at lower metallicities (-0.06 dex and -0.21, respectively), the latter showing a wider distribution in $\rm [Fe/H]$. In order to explain these discrepancies, we explore the age-metallicity relation and the evolution of the fraction of iron produced by Type Ia SNe (Figs. \ref{Fig_AgeMet_1infall} and \ref{Fig_Sigmat_1infall}, respectively). As can be seen in Fig. \ref{Fig_AgeMet_1infall}, the CLOSE G05, MVP06, MR01 and T08 DTDs increase the metallicity up to solar values during the first $\sim$2~Gyr to continue afterwards with in a more steady evolution up to $\rm [Fe/H]\approx0.25$~dex. As Fig. \ref{Fig_Sigmat_1infall} shows, within the initial $1.4$~Gyr the Type Ia SN explosion becomes the dominant iron producing mechanism for the mentioned DTDs, especially for the CLOSE G05 DTD. On the contrary, this transition occurs $2$~Gyr later for the S05 and P08 DTDs. The WIDE G05 corresponds to an intermediate case, showing a more quenched iron production after the first $\sim 2$~Gyr. The age-metallicity relations for the WIDE G05 and P08 DTDs can explain the peaks at lower metallicities in their MDFs: since in these scenarios the synthesis of iron is slower, most of stars are formed at lower metallicities compared to the other DTDs, requiring more time to reach the plateau value.
\par The S05 DTD presents the more complex age-metallicity relation, with three different regimes: during the first $2.5$~Gyr the IRA mechanism drives the production of iron up to $\rm [Fe/H]\approx-0.6$~dex. At this metallicity, the $\rm [Fe/H]$ vs. age curve flattens contributing to the peak observed in the MDF. At later times, the Type Ia SNe accelerate the synthesis of iron during the next $3.5$~Gyr to the saturation at $\rm [Fe/H]\approx0.25$~dex, showing a more extended plateau compared to the other DTDs.
%
%
%
%
%
%
%
%
\section{Galactic Archaeology with the analytic solution: the disc bimodality in the chemical space}
\label{sect_bimo}
\begin{figure*}
\begin{centering}
\includegraphics[width=0.98\textwidth]{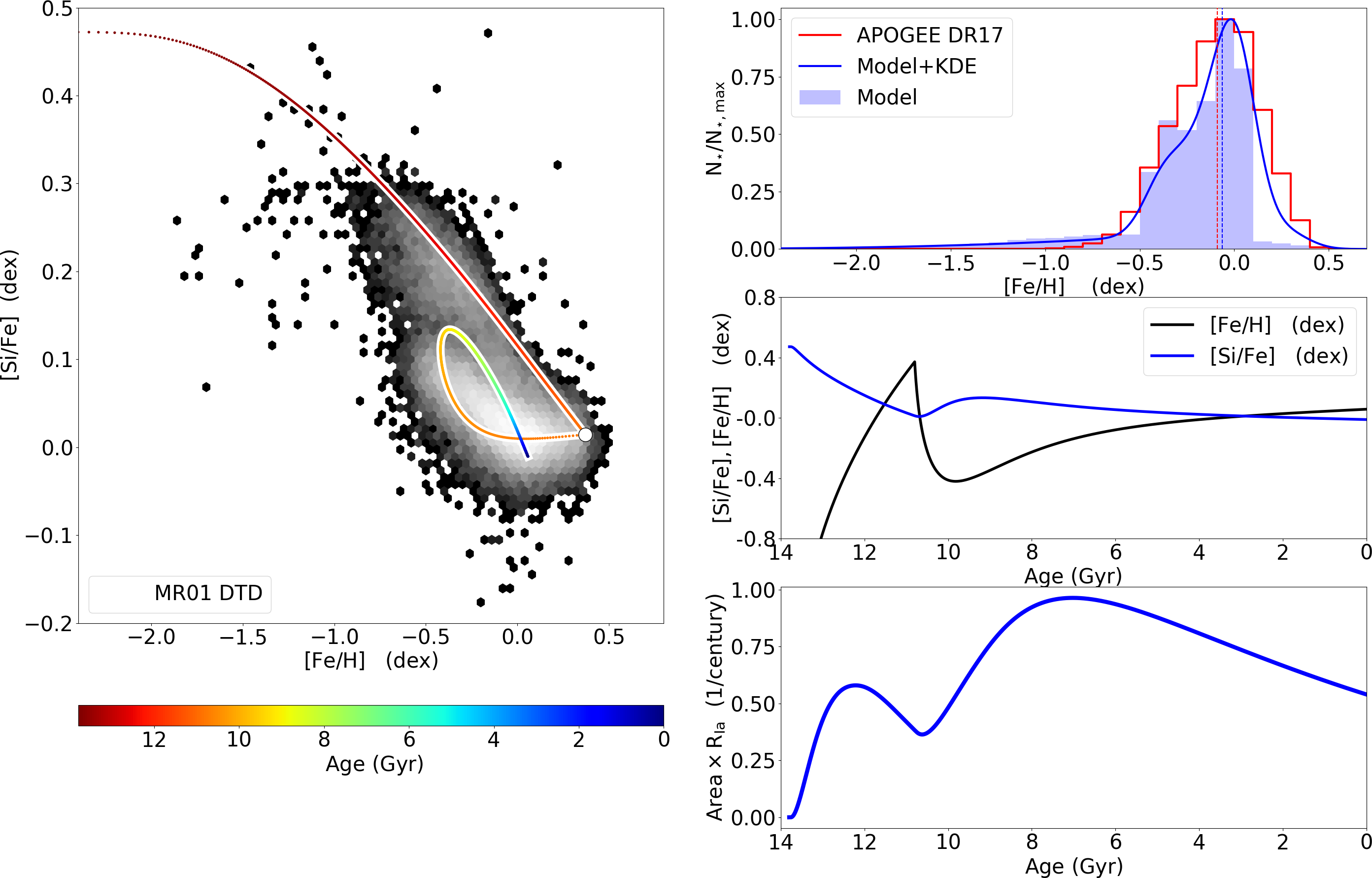}
\caption{{\it Left panel}: Comparison of the [Si/Fe] versus [Fe/H] predicted by our two-infall model in the solar neighbourhood (see model details in Section \ref{sect_bimo}) with the observational data of APOGEE DR17 \citep{apogeedr172022} in the Galactic region between 7.2 and 9.2 kpc. The colour-code throughout the model track stands for the ages of the SSPs formed during the Galactic disc evolution. The white circle represents the  [Si/Fe] versus [Fe/H] value at the beginning of the second infall ($t_G-t_2=10.8$~Gyr). 
{\it Right Upper panel}: MDF expected from the two-infall model shown in Fig. \ref{Fig_ChemSpace_2infall} (blue histogram) compared to the MDF of the APOGEE DR17 sample (red empty histogram). The vertical lines indicate the median values of each distribution. 
{\it Right Middle panel}: The age-metallicity (black line) and [Si/Fe] versus age (blue line) relations predicted by the same model as the upper panel. {\it Right Lower panel}: The temporal evolution of the Type Ia SN rate for the whole Galactic disc is indicated with the blue line.} 
\label{Fig_ChemSpace_2infall}
\end{centering}
\end{figure*}
\begin{figure}
    \centering
    \includegraphics[width=0.48\textwidth]{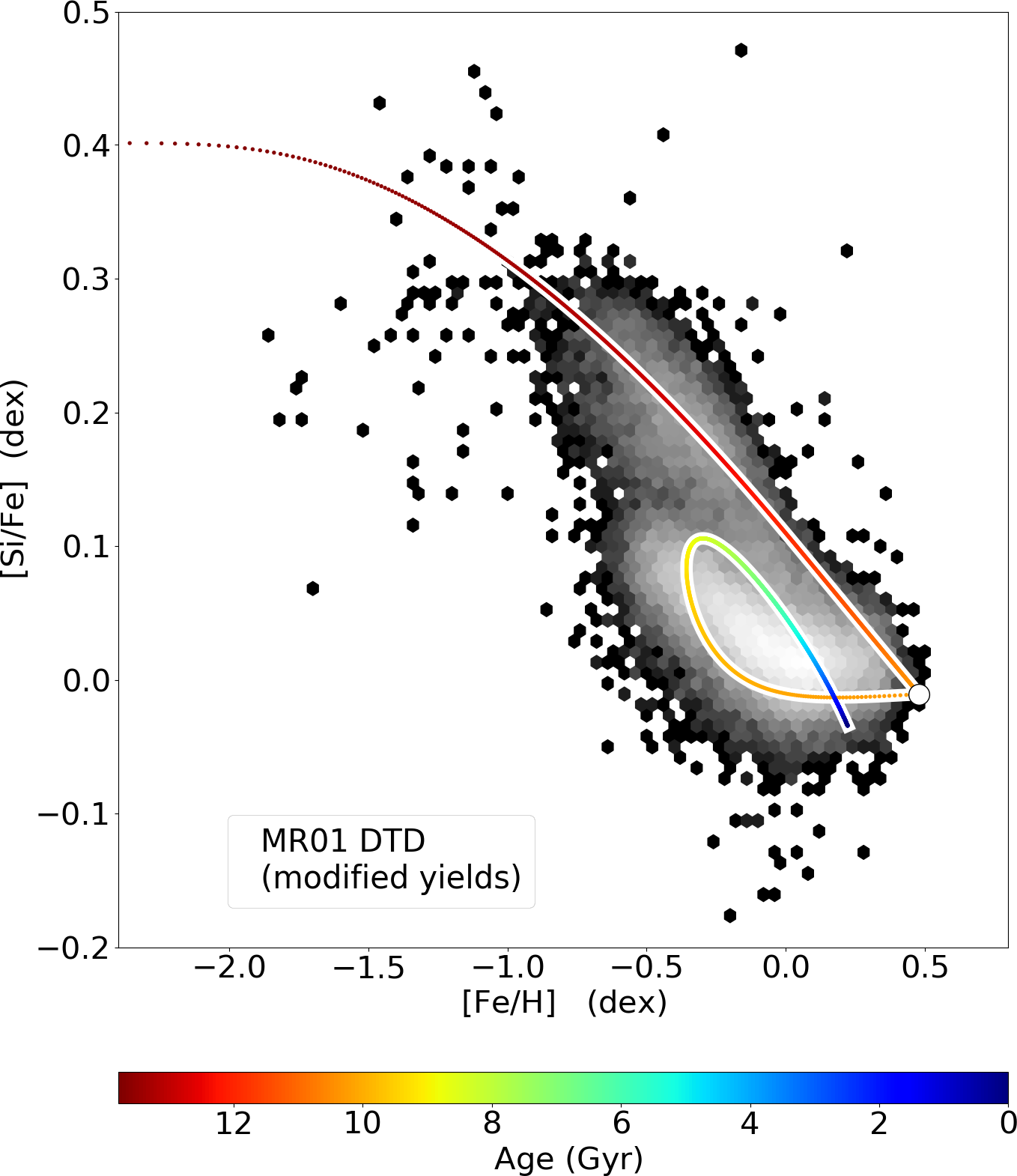}
    \caption{  The same as left panel in Fig. \ref{Fig_ChemSpace_2infall} but considering an alternative model with a smaller loading factor for the wind ($\omega=0.2$), reducing both $\langle y_{\textnormal{Si}}\rangle$ and  $\langle m_{\textnormal{Si},Ia}\rangle$ for silicon by a factor of 15\%.}
    \label{Fig_ChemSpace_2infall_specialyields}
\end{figure}
\begin{figure*}
    \centering
    \includegraphics[width=0.99\textwidth]{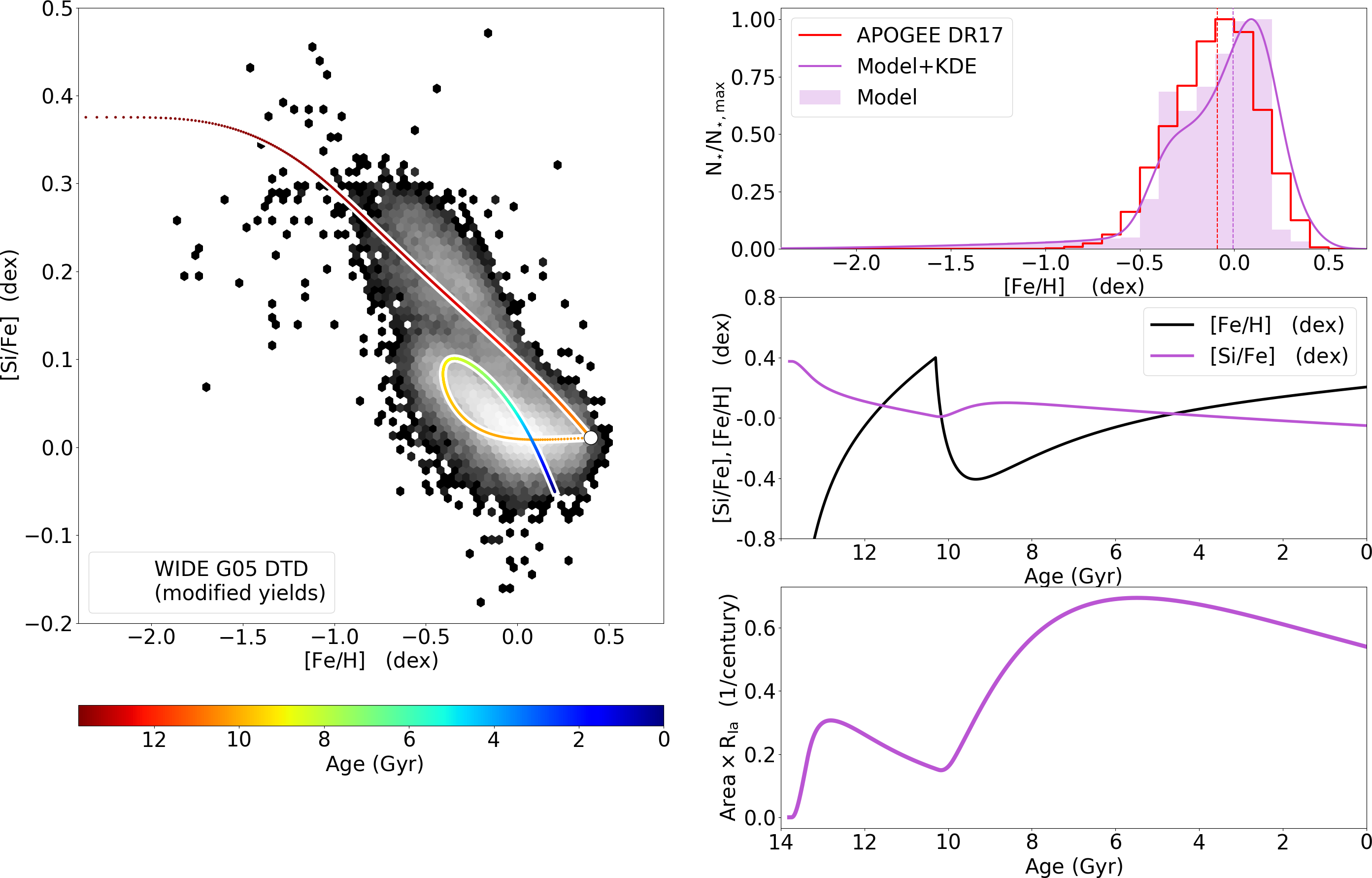}
    \caption{Similar to Fig. \ref{Fig_ChemSpace_2infall} but assuming the WIDE G05 DTD. The yields $\langle y_{\textnormal{Si}}\rangle$ and $\langle m_{\textnormal{Si},Ia}\rangle$ for silicon have been re-scaled by a factor of 80\%.}
    \label{Fig_ChemSpace_2infall_specialyields_WideG05}
\end{figure*}
\par From the chemical point of view, the Galactic disc shows two substructures in the [$\alpha$/Fe] vs [Fe/H] plane: the so-called high-$\alpha$ sequence, classically associated with an old population of stars (thick disc), and the low-$\alpha$ sequence, characterised by the younger stars of the thin disc \citep{fuhrmann2004, reddy2006,bensby2014, Lee11, haywood2013, adibekyan2013}. This dichotomy has been confirmed by the analysis of APOGEE data \citep{Nidever:2014fj, hayden2015,Ahumada2019,queiroz2020,apogeedr172022}, the Gaia-ESO survey \citep[e.g.,][]{RecioBlanco:2014dd,RojasArriagada:2016eq,RojasArriagada:2017ka}, AMBRE \citep{Mikolaitis:2017gd,santosperal2021}, GALAH \citep{buder2019, buder2021}, LAMOST \citep{zheng2021} and Gaia DR3 \citep{recioDR32022b,recioDR32022a}.
\begin{figure*}
    \centering
    \includegraphics[width=0.99\textwidth]{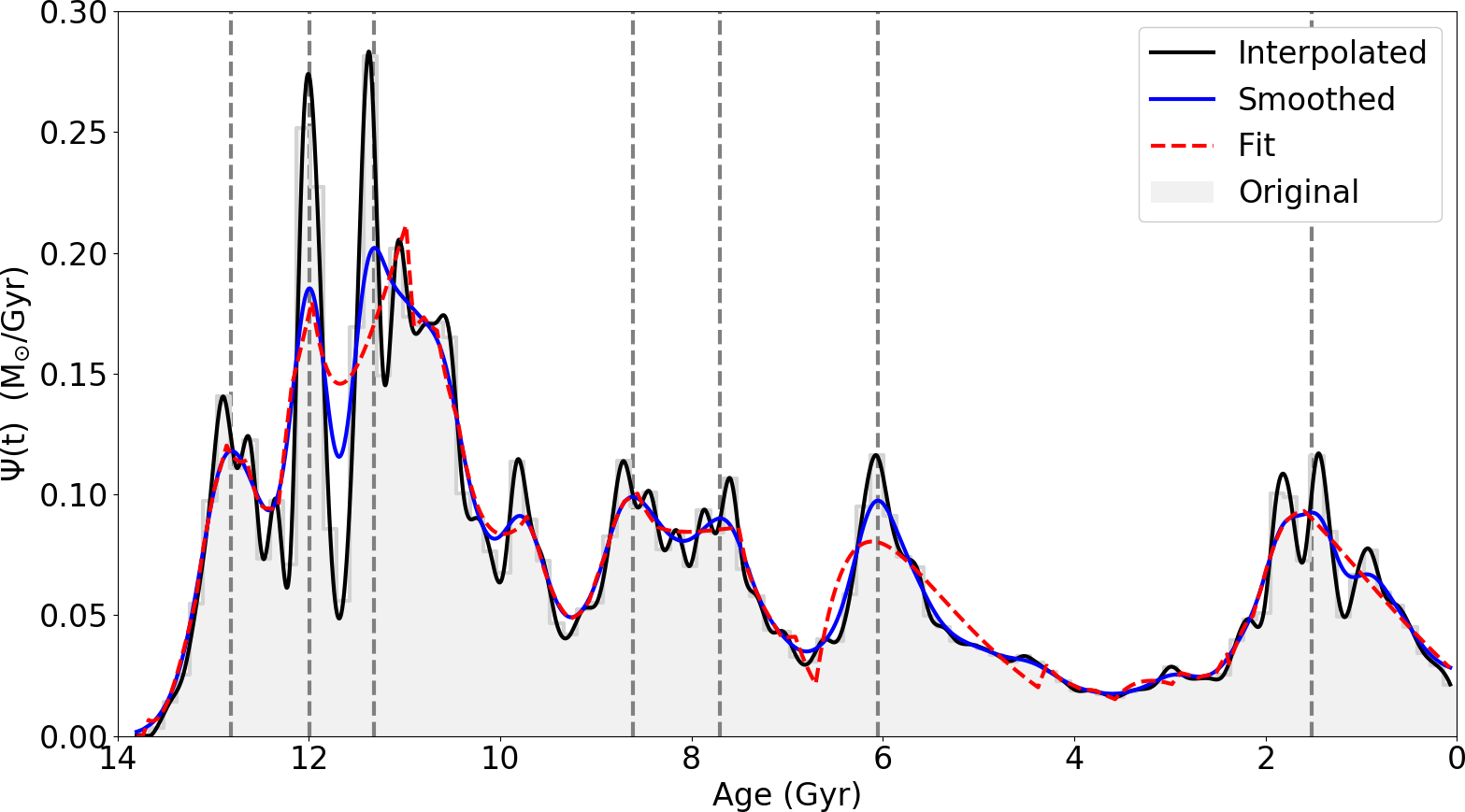}
    \caption{SFR of the \textsc{Galactica} simulated galaxy explored in \citet{park2021} normalised to a total mass of $1 M_{\odot}$. The grey histogram represents the reported SFR, while the solid blue curve corresponds to the smoothed version of the interpolated curve (solid black line). The fit of the smoothed curve is represented by the dashed red line. The positions of the maxima $t_{max}$ used in the determination of $t_k$ (see the text) are indicated by the vertical dashed lines. }
    \label{fig:psi_peirani}
\end{figure*}
\citet{spitoni2019} and \citet{spitoni2021} revise the classical two-infall chemical evolution model \citep{chiappini1997} to reproduce the trends in the [$\alpha$/Fe] vs [Fe/H] diagram observed in the APOKASC  \citep{victor2018} and APOGEE DR16 \citep{Ahumada2019} samples, respectively. In this Galactic formation scenario, the thick and thin disc components have been formed by to distinct episodes of gas accretion. The authors suggest the presence of a  $\sim$~4~Gyr time delay between these episodes in order to reproduce the high- and low-$\alpha$ sequences, imposing precise asteroseismic ages as a constraint.
\par Here, we apply the new analytic solution introduced in Section \ref{sec_results}  in the framework of the two-infall model in order to reproduce the new APOGEE DR17 data \citep{apogeedr172022} for the abundance ratio [Si/Fe] versus [Fe/H] in the annular region 7.2~kpc$<R<$~9.2~kpc (i.e., $R_{\odot}\pm 1$~kpc). As in \citet{spitoni2021}, we impose a signal-to-noise ratio (S/N)>80, a surface gravity log g < 3.5 and vertical height $|Z|<1$ kpc in our selection.
\par Using the same formalism and notation introduced in Section \ref{sec_che_eq}, this infall rate can be written as:
\begin{equation}
\mathcal{I}(t)= \overbrace{ A_1 \,\theta(t)\,  e^{-t/ \tau_{1}}}^{\text{{1st infall, high-$\alpha$}}}+ 
 \overbrace{A_2 \,\theta(\Delta t_2) \,  e^{-\Delta t_2/ \tau_{2}}}^{\text{{2nd infall, low-$\alpha$}}}.
 \label{eq_2infall}
 \end{equation}
\par We impose the present total surface mass density (sum of high- and low $\alpha$ sequence contributions) of $\sigma_{tot}(t_G)=47.1 \pm$  3.4 M$_{\odot} \mbox{ pc}^{-2}$ suggested by \citet{mckee2015} for the local disc. 
\par Initially, we evaluate our analytic solution for the two infall model by assuming the MR01 scenario for the DTD, in which the parameters of the infall are adapted to mimic the observed [Si/Fe] vs. [Fe/H]. For the MR01 DTD, these parameters are $\tau_1=0.4$~Gyr, $\tau_2=7.0$~Gyr, $t_1=0$~Gyr, $t_2=3$~Gyr, $A_1\approx35.128$~$\rm M_{\odot}\,pc^{-2}\, Gyr^{-1}$ and $A_2\approx10.207$~$\rm M_{\odot}\,pc^{-2}\, Gyr^{-1}$. This combination of $A_1$ and $A_2$ implies a second infall four times more massive than the first one. The star formation efficiency $\nu_L$ is set to 0.75~$\rm Gyr^{-1}$ and the loading factor for the wind is $\omega=0.8$.
\par As the left panel of Fig. \ref{Fig_ChemSpace_2infall} illustrates, we can recover the characteristic "loop" in the low-$\alpha$ sequence already found in the detailed chemical evolution models with delayed gas infalls of \citet{calura2009, spitoni2019,palla2020,romano2020,cescutti2022}. This delayed infall creates the low-$\alpha$ sequence by bringing pristine metal-poor gas into the system, which dilutes the metallicity of interstellar medium while keeping [$\alpha$/Fe] abundance almost unchanged. When star formation resumes, the Type II SNe produce a steep increment in the [$\alpha$/Fe] ratio. At later times,  the pollution from the Type Ia SNe raises the metallicity and decreases [$\alpha$/Fe]. This sequence creates a loop in the [$\alpha$/Fe] versus [Fe/H] diagram that overlaps with the region spanned by the APOGEE DR17 \citep{apogeedr172022} data. 
\par It is important to underline that in the \citet{spitoni2021} model no Galactic winds have been considered to fit the APOGEE DR16 data. On the contrary, we impose a significant mass loss due to the Galactic winds ($\omega$=0.8) to reproduce the APOGEE DR17 data. Possibly it is due to the nucleosynthetic prescriptions for massive stars used in that work, which, in line with \citet{francois2004}, include a modification of the \citet{WW1995} yields in order to mimic the data available in the solar vicinity. Motivated by this explanation, we multiply $\langle y_{\textnormal{Si}}\rangle$ and $\langle m_{\textnormal{Si},Ia}\rangle$ by a factor of $0.85$ to reproduce the [Si/Fe] vs. [Fe/H] diagram observed with APOGEE data (Figure \ref{Fig_ChemSpace_2infall_specialyields}). Using these yields, we can model the chemical evolution track with a lower wind loading factor ($\omega=0.2$), while for the rest of the parameters we consider $\tau_1=0.13$~Gyr, $\nu_L=0.8$~Gyr$^{-1}$, $\tau_2=6.75$~Gyr , $t_2=3.5$~Gyr, $A_1\approx52.886$~$\rm M_{\odot}\,pc^{-2}\, Gyr^{-1}$ and $A_2\approx6.898$~$\rm M_{\odot}\,pc^{-2}\, Gyr^{-1}$ (mass ratio between the two infalls of $5.3$). 
\par In Fig. \ref{Fig_ChemSpace_2infall}, we show the age-metallicity relation predicted by our model, where the effects of the dilution produced by the delayed infall is clear. The consequent gap in the star formation rate has a significant effect on the Type Ia SN rate (local minimum at age $\approx10.5-10.0$~Gyr in the lower right panel in Fig. \ref{Fig_ChemSpace_2infall}). A similar feature in the age-metallicity relation has been found by \citet{nissen2020} in the analysis of the HARPS spectra of local solar-like stars. They note that the distribution of stars in the age-metallicity relation has two distinct populations with a clear age dissection. The authors suggest these two sequences may be interpreted as an evidence of two gas accretion episodes onto the Galactic disc, with a quenched star formation between them. This is in agreement with the scenario proposed by \citet{spitoni2019} and with the results shown here. By analysing subgiant stars of LAMOST, \citet{xiang2022} identify two distinct sequences in the stellar age-metallicity distribution separated at age $\sim$~8~Gyr. Similarly, \citet{sahlholdt2022} propose an age-metallicity relation characterised by several disconnected structures, which could be linked to different star-formation regimes throughout the Milky Way disc evolution.
\par Figure \ref{Fig_ChemSpace_2infall} shows the comparison of the MDFs predicted by our two-infall model and that observed in the APOGEE DR17 data. We can see that, although both distributions have similar median values, their shapes differ. This difference between the predicted and the observed MDFs is more significant in the super metal-rich regime \citep[\textnormal{$\rm \lbrack Fe/H\rbrack\gtrsim0.1$~dex,}][]{santosperal2021}, where our two-infall model sub-estimates the number of sources at that metallicity. This discrepancy can be explained by the effect of the radial migration from the inner Galaxy: stars born in the central high-metallicity regions have experienced a change in their angular momentum due to the interaction with the non-axisymmetric structures of the disc, like the bar and the spiral arms \citep{Sellwood2002,SchBinney09, Minchev11}. Such migrated population shapes the metal-rich tail of the MDF, increasing its skewness as reported by \citet{hayden2015} and laterly confirmed by \citet{Loebman2016, MartinezMedina16, MartinezMedina17}. Since Eq. \ref{Eq_X_Fe} does not include any term associated with the radial migration, our analytic solution predicts a lower number of super-solar metallicity stars. However, by blurring the distribution using a Gaussian Kernel Density Estimator of width 0.1 dex, we obtain a smooth distribution whose shape agrees better with that of the APOGEE MDF in the super-solar regime.
%
%
%
%
\par We evaluate the dependence of the two-infall chemical evolution model on the DTD by repeating the previous analysis with the WIDE G05 DTD. We discarded the use of the CLOSE G05 DTD for this test because no good combination of infall parameters has been found. In short, this DTD provides a large fraction of prompt events, so that the Fe enrichment occurs very fast and the MDF results overpopulated at high metallicities for all the realistic options of infall parameters tested (see Appendix \ref{Sec_CLOSEG05}). Similarly, we use a different set of parameters compared to the MR01 case because no satisfactory common parameters have been found. For the WIDE G05 DTD, the parameters of the infall that better reproduce the APOGEE DR17 data are $\tau_1=0.13$~Gyr, $\tau_2=6.75$~Gyr, $t_1=0$~Gyr, $t_2=3.5$~Gyr, $A_1\approx52.887$~$\rm M_{\odot}\,pc^{-2}\, Gyr^{-1}$ and $A_2\approx6.898$~$\rm M_{\odot}\,pc^{-2}\, Gyr^{-1}$ (also equivalent to a mass ratio between infalls of $5.3$), while the star-formation efficiency and the wind loading factor have been set to $\nu_L=0.8$~Gyr$^{-1}$ and $\omega=0.2$, respectively. As in the MR01 case, we consider a rescaled version of the silicon yields by applying a factor $0.8$ to the nominal values presented in Section \ref{sec_1infall}. As we can see in Fig. \ref{Fig_ChemSpace_2infall_specialyields_WideG05}, the resulting chemical evolution track is able to reproduce the observed $\rm [Si/Fe]$ vs. $\rm [Fe/H]$. In the high-$\alpha$ regime, the solution with the WIDE G05 DTD shows a similar trend to that found with the MR01 DTD, while for the low-$\alpha$ sequence it requires an earlier second infall and a more extended tail loop to trace the chemical evolution. Compared to Fig. \ref{Fig_ChemSpace_2infall}, the model with the WIDE G05 DTD results in a more metal rich MDF, with a peak at $\rm [Fe/H]\approx 0.09$~dex and a larger discrepancy with the median metallicity of the APOGEE DR17 sample ($\Delta \rm[Fe/H]=0.08$~dex).
%
%
%
%
%
%
\section{The Milky Way-like disc in the cosmological context}
\label{Sec_Cosmo}
\par Numerical simulations constitute an important tool for the study of the formation and evolution of galaxies \citep{vogel2020}. They allow for the comparison of the structure, kinematics and chemical composition inferred from the observational data with the models. Some simulations, however, are limited by their lack of chemical information. We can overcome this restriction by modelling the chemistry in this simulations with the analytic solution presented in this work.
\par In this section, we propose to apply our analytical model to one simulated Milky-Way like galaxy, nicknamed \textsc{Galactica}, which was introduced in \citet{park2021}. \textsc{Galactica} is extracted from a zoom-in hydrodynamical simulation in a cosmological context (i.e. the region of interest, including the galaxy and its host dark matter halo, has been re-simulated at much higher resolution), using the same spatial resolution ($\sim$40~pc) and
the same sub-grid models than the NewHorizon simulation \citep[see ][]{Dubois2021}. Using the SFH of \textsc{Galactica}, we aim at modeling the chemistry of that Galaxy by applying our analytic solution to Eq. \ref{eq_to_solve}. Figure \ref{fig:psi_peirani}) shows the SFR of \textsc{Galactica} \citep[same as in Fig.10 of ][]{park2021} normalised to a total integrated mass of $1 M_{\odot}$ and re-scaled in time to set its age at $t_G=13.8$~Gyr (3\% older than in the original work).
\par In order to fix some fitting problems, we oversample the binned SFR using a third order spline interpolator evaluated in a grid of time nodes of step size $\Delta t=0.005$~Gyr (black curve in Fig. \ref{fig:psi_peirani}). We smooth this curve by performing a convolution with a Gaussian kernel of width $0.15$~Gyr. The resulting SFR is multiplied by a renormalisation factor to keep the total mass of $1 M_{\odot}$ fixed (blue curve). As Fig. \ref{fig:psi_peirani} shows, this process redistributes the stellar mass near the most prominent peaks and leads to a less spiky stellar formation history.
\par By analogy with Eq. \ref{eq:sfr_sol}, we propose a fit for the smoothed SFR by a superposition of functions $\psi_k (t)$ of the form
\begin{equation}
    \label{Eq_func_basis}
    \psi_k(t) = C_k\cdot \theta(t-t_k)\left[\exp\left(-\dfrac{(t-t_k)}{\tau_k}\right)-\exp\bigg(-\alpha(t-t_k)\bigg)\right]   
\end{equation}
where the free parameters are the amplitudes $C_k$, the timescales $\tau_k$ and the offsets $t_k$. The parameter $\alpha$ is given as input and set to 2.23~Gyr$^{-1}$ (equivalent to considering $\nu_L=2$~Gyr$^{-1}$ and $\omega=0.4$). In contrast to Eq. \ref{eq_2infall}, we do not include the $\sim \exp(-\alpha t)$ term because we find better a fit without it. Among the free parameters, only the amplitudes can be determined exactly by the linear least-squares fitting method. On the contrary, the timescales and offsets require more advanced optimisation techniques whose convergence can be very slow. For this reason, we propose the following alternative for $\tau_k$ and $t_k$:
\begin{itemize}
\item We select the position of the local maxima of the interpolated SFR indicated ($t_{max}$) in Fig. \ref{fig:psi_peirani} (vertical dashed lines).
\item We consider the set of timescales from $\tau_k=1$~Gyr to $\tau_k=5$~Gyr (step size of $1$~Gyr) adding $\tau_k=8, 10$~Gyr.
\item Differentiating both sides of Eq. \ref{Eq_func_basis}, we find the maximum of $\psi_k (t)$ is located at $t_{max}=t_k + \tau_k\ln{(\alpha \tau_k)}/(\alpha\tau_k-1)$. Thus, we solve for $t_k$ for all the possible combinations ($\tau_k$, $t_{max}$). If the resulting offset is negative, we substitute it by the half of the minimum positive $t_k$.
\item We construct a basis of $\psi_k$ functions with all the combinations of ($\tau_n$, $t_k$), but excluding the cases in which both $\tau_n$ and $t_{max}$ are larger than $6$~Gyr. This results in a set of 60 functions $\psi_k(t)$.
\item Finally, using the standard least-squares fitting method we compute the values of the amplitudes $C_k$.
\end{itemize}
\begin{figure}
    \centering
    \includegraphics[width=0.48\textwidth]{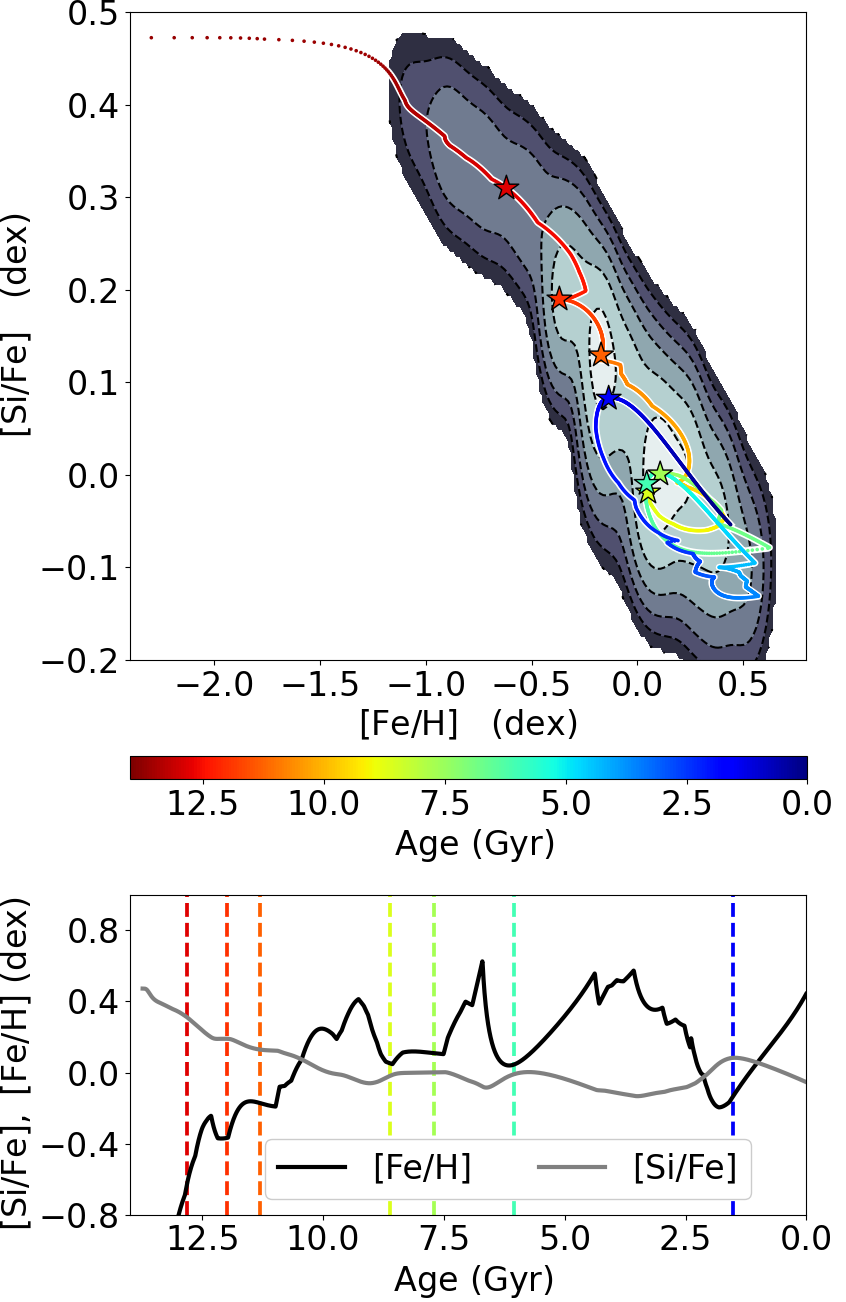}
    \caption{ {\it Upper Panel:} [Si/Fe] versus [Fe/H] extracted from the star formation history shown in Fig. \ref{fig:psi_peirani}. The evolution track is colour-coded with the stellar ages while the star markers indicate the maxima in the SFR ($t_{max}$). The evolution track is blurred with a 2D Gaussian kernel of width 0.05 dex (grayscale contour plot).
    {\it Lower Panel:} The age-metallicity (solid black line) and [Si/Fe] versus age (solid gray line) relations predicted by the same model as the upper panel. Dashed vertical lines denote the ages associated with $\rm t_{max}$ (star makers in the upper panel and dashed vertical lines in Fig. \ref{fig:psi_peirani}) and follow the same colour-code convention as the upper panel.}
    \label{fig:alpha_peirani}
\end{figure}
\par The resulting fitting function is illustrated in Fig. \ref{fig:psi_peirani} (dashed red curve). Although some deviations are observed, it traces the general trend of the smoothed SFR. In order to get realistic chemical evolution tracks, we multiply the SFR by a factor to get the present-day stellar density of $\sigma_{\star}(t_G)=33.4$ $\pm$ 3 M$_{\odot} \mbox{ pc}^{-2}$ \citep{mckee2015}.
\par In Fig. \ref{fig:alpha_peirani}, we show the [Si/Fe] versus [Fe/H] abundance ratios predicted by our analytic model from the fitting SFR mentioned above. We note that for ages older than 10 Gyr, the SFH traced by the fitting function shows three peaks without significant extended quenching periods between them. Their imprint on the [Si/Fe] versus [Fe/H] is characterised by a ``smooth'' evolution, with a mild dilution signature associated with the second peak of star formation. Nevertheless, some burst features can be found, as already discussed in the analysis of the high-$\alpha$ sequence of  the Milky Way-like zoom-in cosmological   {\tt  VINTERGATAN} simulation \citet{agertz2021}. At more recent ages, the subsequent infalls of pristine gas produce a depletion of [Fe/H], especially in the age intervals $7-6$~Gyr and $3-2$~Gyr, while it increases during the following extended periods of low star formation activity. As expected, the ``loop'' features become more prominent and extended at recent times, creating a low-$\alpha$ structure at super-solar metallicities. 
\par It is worthwhile mentioning that our analytic solution has been computed imposing a constant value for the star formation efficiency $\nu_L$. As discussed in \citet{spitoni2022}, two chemical evolution models constrained by the same SFH can lead to different enrichment of silicon and iron just imposing a less massive gas infall and a higher star formation efficiency. In their Fig. 5, they show that for this case the dilution is substantially diminished. Thus, with the presented analytic solution, we maximise the dilution effect (through a massive infall of pristine gas) since $\nu_L$ cannot increase during the Galactic evolution nor mimic the strong star burst phases.
\par We have checked that in the \textsc{Galactica} simulation the peaks of the star formation (Fig. \ref{fig:psi_peirani}) are associated with a rapid increase of the gas mass. Therefore, the scenario proposed by our model, in which the SFH is the result of 60 subsequent events of gas infall, is valid. In any case, according to several chemical evolution models \citep{spitoni2019,palla2020,lian2020} and chemo-dynamical simulations \citep{agertz2021,khoperskov2021,vincenzo2020}, the dilution effect originated by the accretion of pristine (or mildly chemical enriched) gas should dominate the chemical enrichment of the low-$\alpha$ sequence.
%
%

\section{Conclusions and Future work}
\label{concl}
In this work we present a new analytic solution to the Galactic chemical evolution model which can be used with different prescriptions of the DTD, including the single and double degenerate scenarios. We provide some examples of possible applications of our solution, whose main conclusions are summarised as follows:
\begin{itemize}
    \item  We prove that our solution can constitute a useful tool for Galactic Archaeology by interpreting the chemical  APOGEE DR17 disc stars.
    The analytic solution can reproduce the expected chemical evolution of the $\alpha$ and iron-peak elements. In particular, we compare the pattern in the [Si/Fe] vs. [Fe/H] plane observed by APOGEE DR17 with these predicted by two different models: one assuming a Single Degenerate DTD and another that considers the Double Degenerate scenario. In both cases, we find the low-$\alpha$ sequence can be explained by a delayed gas infall, in  agreement with the results of detailed numerical models, but considering different Galactic and infall parameters.
    \item The super-solar metallicity regime observed in APOGEE DR17 is poorly reproduced by our solution since the considered chemical evolution model does not include radial migration terms. However, the blur of the predicted MDF with a Gaussian Kernel of width $0.1$~dex improves the comparison.
    \item According to our tests, it is not possible to discern the best DTD for reproducing the data. With the suitable realistic combination of parameters, both the MR01 and the WIDE G05 DTDs can predict the two sequence pattern seen in the $\rm [Si/Fe]$ diagram, as well as the approximated shape for the MDF. In order to break this degeneracy, more constraints based on accurate stellar ages, more precise stellar yields and gas infall timing among others are required.
    \item  By modelling the chemistry of a simulated Milky Way-like galaxy from its star formation history, we exploit the applicability of our solution in a cosmological context. The study presented here for the \textsc{Galactica} simulation constitutes a preliminary work which will be extended with galaxies of different morphology and formation history.
\end{itemize}
\par In future, we plan to include in our solution the contribution of periodic perturbations, such as these caused by the spiral arms \citep{spitoni2D2019,poggio2022,palicio2022} and bars \citep{palicio2018}. We also aim to extend our analytic solution to two-dimensional and three-dimensional models by including gas flows, transport of metals as well as radial migration.
\par Since the analytic solution presented here can be used to model dwarf galaxies, it is possible to perform Bayesian fits of Local Group galaxies with our solution \citep{johnson2022}. Similarly, the chemistry of galaxies with different morphology can be addressed. For instance, for early-type galaxies it should be possible to compare the predictions for $\langle$[$\alpha/\text{Fe}$]$\rangle$ with the results of the  MaNGA survey \citep{liu2020}.
Moreover, it will possible to characterise  the star-forming objects which obey to scaling-relations, like the main sequence star formation \citep{spitoni_MZ2020,spitoni2021MDF} providing for them the [$\alpha$/Fe] evolution in time.
%
%
%
%
%
%
%

\begin{acknowledgements}
 We want to thank the helpful comments from P. de Laverny. P. A. Palicio acknowledges the financial support from the Centre national d'études spatiales (CNES). E. Spitoni and A. Recio-Blanco received funding from the European Union’s Horizon 2020 research and innovation program under SPACE-H2020 grant agreement number 101004214 (EXPLORE project). 
\end{acknowledgements}

%
%
%
\bibliographystyle{aa}  
\bibliography{biblio} 
\begin{appendix}
\section{Fitting procedure of the DTDs}
\label{app_DTD_technical_section}
\par As mentioned in Section \ref{DTD_sec}, the DTDs whose functional form cannot be exactly described by Eq. \ref{Eq_DTD_component} require a fitting approximation in order to be evaluated in the analytic solutions. In this section, we detail the procedure performed for the individual modelling of such DTDs and evaluate their associated errors.
\subsection{The case of the MR01 DTD}
\label{DTD_MR01_Sec}
\par According to MR01 and Fig. \ref{Fig_DTDs}, the MR01 DTD is described by a piece-wise function of two components connected at $t_{\rm knee}\approx 1.612$~Gyr whose shape and slope depends on the $\gamma$ parameter, in which we assume $\gamma=0.5$ as in \citet{bonaparte2013}. We model the leftmost component as follows:
\begin{itemize}
    \item Initially, we perform a Gaussian Mixture fitting assuming $N_G$ Gaussian distributions \citep{astroML, astroMLText}. As a result, we obtain three $N_G$-dimensional arrays with the mean values ($\vec{\tau}'$), the widths ($\vec{\sigma}'$) and the amplitudes ($\vec{A}_G$) of the Gaussian curves. We substitute the lowest value in $\vec{\tau}'$ by 0.094~Gyr because we find a better fit of the MR01 DTD.
    \item We propose a set of $N_E$ decreasing exponential curves whose characteristic timescales ($\vec{\tau}_D$) are multiple of a fundamental timescale $\tau_f$ (i.e, the n-th component of $\vec{\tau_D}$ is $n\cdot \tau_f$, with $1\leq n\leq N_E$). The choice for $\tau_f$ is motivated by the naïve fitting of the modal and knee points (A and B of Fig. \ref{Fig_DTDs}) with a single exponential curve, resulting in a timescale of $0.57$~Gyr. We divide this value by a factor of two to account for shorter timescales; thus, $\tau_f\approx0.57$~Gyr.
    \item We add a $\sim t^{-1}$ term to the fitting function. 
    \item Fixing the non-linear parameters $\vec{\tau}'$, $\vec{\sigma'}$ and $\vec{\tau}_D$, we search the values of the $N_G~+~N_E~+~1$ amplitudes ($A_{G,i}$, $A_{E,i}$ and $A_{I,i}$, respectively) that minimise the discrepancy with the MR01 DTD. By using Lagrange multipliers, we impose four additional constraints on the least-squares fitting algorithm: we fix the values of the fit at $t=0.03$~Gyr, $\sim0.094$~Gyr (the maximum) and $1.612$~Gyr (the ``knee''), and impose zero derivative at the maximum.
    \item We repeat this procedure testing different combinations of $N_G$ and $N_E$ to find a good compromise between the complexity of the fitting and the similarity with the MR01 DTD. Based on these tests, we select the combination $N_G=3$, $N_E=5$ (see the first five rows in Table \ref{tab_dtd}).
\end{itemize}
\par For the rightmost part of the MR01 DTD ($t>t_{\rm knee}$), we repeat the previous procedure with the following modifications:
\begin{itemize}
    \item No imposed values are used for $\vec{\tau}'$.
    \item The fundamental timescale $\tau_f$ is computed using the coordinates of the ``knee'' and the minimun (located at $t=13.8$~Gyr), which leads to $\tau_f\approx 1.79$~Gyr.
    \item Using the restricted least-squares fitting algorithm, we fix the values at the edges of the interval [$t_{knee}$, $13.8$~Gyr].
    \item After testing different combinations, we consider the case with $N_E=4$ and no Gaussian distributions the best choice for this part of the MR01 DTD.
\end{itemize}
\par
\begin{figure}
\begin{centering}
\includegraphics[width=0.45\textwidth]{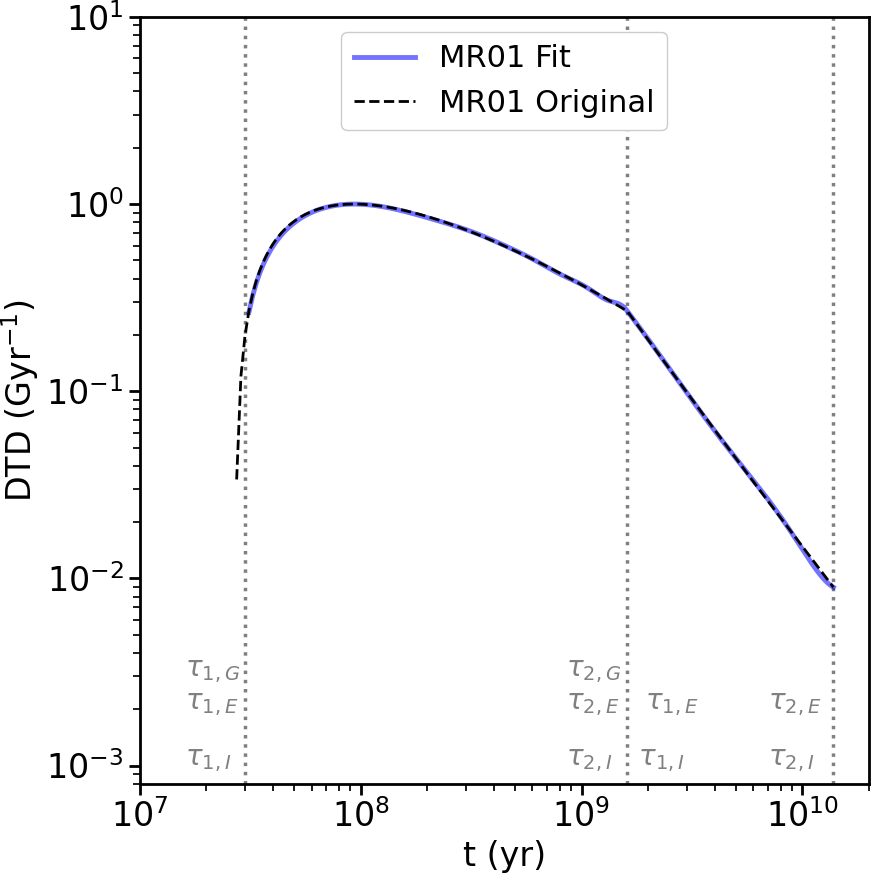}
\caption{Delay time distribution of \citet{matteucci2001} normalised to the maximum (dashed black curve) and its fit using three Gaussian curves, nine exponentials and two $t^{-1}$ functions (solid blue line). Vertical dotted lines indicate the time domain $(\tau_1, \tau_2) _{\text{\,G, E, I}}$ of the fitting for each section of the DTD (see Table \ref{tab_sel} for the adopted parameters).}
\label{Fig_best_fit_MR01}
\end{centering}
\end{figure}

 The resulting set of parameters for this DTD are summarised in the upper part of Table \ref{tab_dtd}. Figure \ref{Fig_best_fit_MR01} illustrates the comparison of the original MR01 DTD with its fit, whose maximum discrepancy in absolute value is 0.018. We note the fitting function is defined on a shorter time interval than that of the MR01 because the former becomes negative when $t\lesssim 30$~Myr, while the MR01 DTD is defined for $t\gtrsim27$~Myr. This interval of $\sim3$~Myr, however, has a negligible contribution to the total area ($\sim0.02$~\%) of the MR01 DTD. 
\begin{figure*}
\begin{centering}
\includegraphics[width=0.935\textwidth]{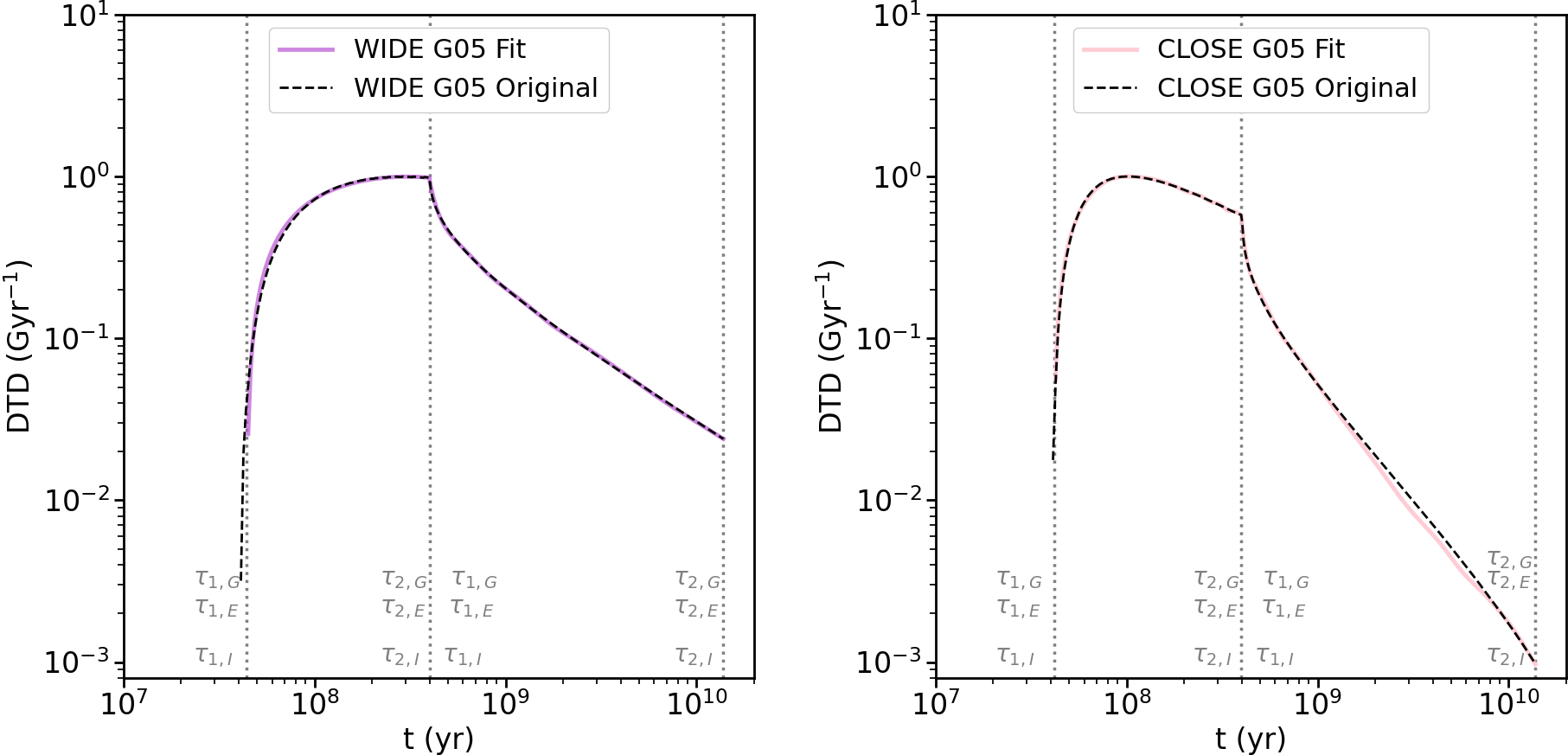}
\caption{WIDE (left panel) and CLOSE (right panel) double degenerate delay time distributions of \citet{greggio2005} normalised to their maximum (dashed black curves). The form of the WIDE G05 DTD is defined by the parameters $\tau_{n,x}=0.4$~Gyr, $\beta_a=0$ while for the CLOSE G05 DTD we use the combination $\tau_{n,x}=0.4$~Gyr, $\beta_g=-0.975$. The WIDE DTD is modeled as six Gaussian curves, eight exponentials and two $t^{-1}$ functions (solid purple line) while the CLOSE DTD requires twelve Gaussian curves, nine exponentials and four $t^{-1}$ functions (solid pink line). Vertical dotted lines indicate the time domain $(\tau_1, \tau_2) _{\text{\,G, E, I}}$ of the fitting for each section of the DTD (see Table \ref{tab_sel} for the adopted parameters).}
\label{Fig_best_fit_G05}
\end{centering}
\end{figure*}
\subsection{The case of the WIDE G05 DTD}
\label{DTD_WideG05_Sec}
\par As Fig. \ref{Fig_DTDs} illustrates, the WIDE G05 DTD increases asymptotically up to its maximum to decrease at later times following a power-law like relation, in which its slope is defined by the $\beta_a$ parameter \citep[see Section 4.3.1 of ][]{greggio2005}. For illustrative purposes, in this work we consider the intermediate case $\beta_a$=0. The transition between the mentioned two regimes is determined by the nuclear timescale of the least massive secondary in Type Ia SN progenitor systems $\tau_{n,x}$, whose value of $0.4$~Gyr adopted in this work implies a mass of 3~$M_{\odot}$. Using $\tau_{n,x}$ as reference, we define two time intervals to perform the fitting.
\par For the $t\leq\tau_{n,x}$ interval, we perform a restricted least-squares fitting procedure similar to those considered for the MR01 DTD. We model the increasing part of the G05 DTD with a Gaussian curve centered at $\tau'=0.09$~Gyr, an exponential function with timescale $\tau_D=3.19$~Gyr and a $1/t$ relation imposing continuity at $\tau_{n,x}$. In contrast, the decreasing regime ($t>\tau_{n,x}$) requires a more complex fitting function: 
\begin{itemize}
    \item We propose a set of five Gaussian curves with $\tau'=\tau_{n,x}$ and widths $\sigma'$ ranging from $0.1$ to $0.5$~Gyr (step $0.1$~Gyr).
    \item For the exponential term, we create a partition of the time interval $[\tau_{n,x}, 13.8\ Gyr]$ into six subintervals and repeat the procedure described in Section \ref{DTD_MR01_Sec} in each subdivision. The resulting timescales $\tau_D$ are summarised in Table \ref{tab_dtd}.
    \item We introduce an offset $\tau_0=0.35$~Gyr in the $1/t$ relation to improve the fitting in the $t \rightarrow\tau_{n,x}$ regime, where the slope of the G05 DTD becomes steeper.
    \item The amplitudes of the twelve fitting functions described above are optimised by the restricted least-squares method fixing the values at $t=\tau_{n,x}$ and $13.8$~Gyr, and the first derivative at $t=13.8$~Gyr.
\end{itemize}
\par The maximum discrepancy between the normalised G05 DTD and its fit is $0.041$ (see right panel in Fig. \ref{Fig_best_fit_G05}). Similarly to the case of the MR01 DTD, we find a difference of $\sim 3$~Myr between the time domains of the fit and the original DTD that excludes a negligible fraction of the WIDE G05 DTD area ($\sim 0.04$\%).
\subsection{The case of the CLOSE G05 DTD}
\label{DTD_CloseG05_Sec}
\par As in the previous cases, we identify two regimes in the CLOSE double degenerate G05 DTD (pink curve in Fig. \ref{Fig_DTDs}) connected at $t=\tau_{n,x}=0.398$~Gyr, where the slope in the power-law regime is determined by the parameter $\beta_g$ \citep[][see Section 4.3.2 of ]{greggio2005}. In this work, we illustrate the case $\beta_g=-0.975$ (i.e., a very steep DTD). The leftmost part of this DTD can be modelled using the following fitting functions:
\begin{itemize}
    \item Six Gaussian curves whose parameters are determined by a Gaussian Mixture process, imposing the value of the lowest offset $\tau'$ to $0.102$~Gyr.
    \item Four exponential curves whose timescales are estimated using the same partition procedure as for the WIDE DTD case.
    \item One  $t^{-1}$ function.
\end{itemize}
The contribution of these eleven functions are optimised by a restricted least-squares algorithm fixing the values at $t=0.04$~Gyr, $0.398$~Gyr and the maximum at $0.102$~Gyr. Similarly, the rightmost part of the CLOSE G05 DTD ($t>0.102$~Gyr) is modelled imposing:
\begin{itemize}
    \item Five exponentials with $\tau_D$ determined by the partition of the [$0.102$~Gyr, $13.8$~Gyr] interval.
    \item Three $(t-\tau_0)^{-1}$ functions with $\tau_0=0.25$, $0.30$ and $0.33$~Gyr.
    \item  Six Gaussian curves whose offsets $\tau'$ and widths $\sigma'$ are updated in an iterative process based on the mismatch between the original CLOSE G05 DTD and its fit. The resulting values are summarised in the third and fourth columns of Table \ref{tab_dtd}.
\end{itemize}
where the constraints fix the values at $t=0.102$~Gyr, $13.8$~Gyr as well as the slope at $t=13.8$~Gyr. This fit results in a maximum discrepancy with the original CLOSE G05 DTD of 0.023 (left panel in Fig. \ref{Fig_best_fit_G05}).
\begin{figure}
\begin{centering}
\includegraphics[width=0.4512\textwidth]{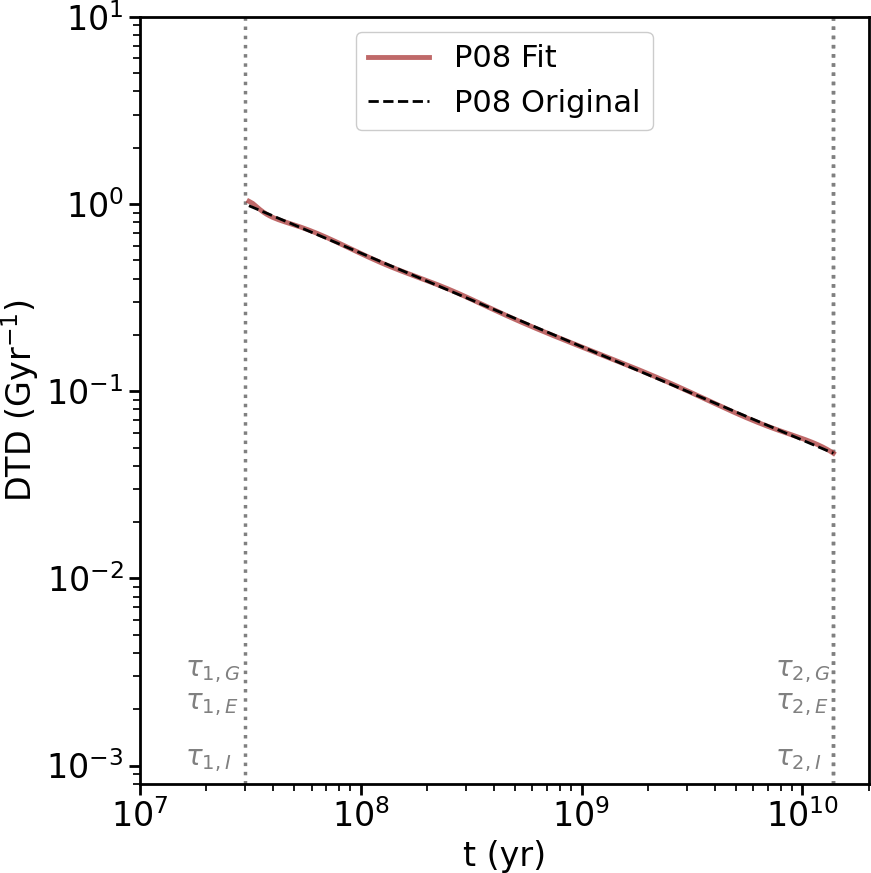}
\caption{Delay time distribution of \citet{Pritchet08} normalised to the maximum (dashed black curve) and its fit using one Gaussian curves, four exponentials and four $(t-\tau_0)^{-1}$ functions (solid brown line). Vertical dotted lines indicate the time domain $(\tau_1, \tau_2) _{\text{\,G, E, I}}$ of the fitting for each section of the DTD (see Table \ref{tab_sel} for the adopted parameters).}
\label{Fig_best_fit_P08}
\end{centering}
\end{figure}
\subsection{The case of the P08 DTD}
\label{DTD_P08_Sec}
\par Since the P08 DTD has the form $\sim t^{-1/2}$ we do not need to use piece-wise functions for its approximation as with the previous DTDs. We model the P08 DTD as a combination of the following functions:
\begin{itemize}
    \item A Gaussian curve with $\tau'=3.5\times 10^{-3}$~Gyr and $\sigma'=0.1$~Gyr.
    \item Four exponential curves whose timescales $\tau_D$ are computed as in the G05 case.
    \item Four $(t-\tau_0)^{-1}$ functions, where $\tau_0\in\lbrace0.025$, $0.02$, $0.015$, $0.01\rbrace$.
\end{itemize}
The amplitudes of these nine basis functions are optimised imposing the exact values at the limits of the fitting interval (see Table \ref{tab_dtd}). This results in a discrepancy between the P08 DTD and its fit lower than 0.017 (Fig. \ref{Fig_best_fit_P08}). 
%
%
%
%
%
\onecolumn
\section{Analytic solution}\label{app_sol}
\subsection{Case $\tau_j\neq\alpha^{-1}$}
In order to simplify the analytic expressions it is useful to define the following parameters and functions:
\begin{eqnarray}
    \label{Parameters}
    \beta_j &=& \alpha-\tau_j^{-1}\\
    \lambda_x(t) &=& \min(t, \tau_{1x})  \ \ \ \ \ \textnormal{with}\ \ x \in \lbrace{G, E, I}\rbrace\\
    \Lambda_x(t) &=& \min(t, \tau_{2x})  \ \ \ \ \ \textnormal{with}\ \ x \in \lbrace{G, E, I}\rbrace\\
    \eta_j &=& \tau'+\sigma'^2 /\tau_j\\
    \eta_\alpha &=& \tau'+\sigma'^2 \alpha
\end{eqnarray}
\begin{eqnarray}
\label{Functions}
    Q_D(x;a|c) &=& \begin{cases}
\dfrac{a \cdot c}{a-c}\cdot\exp{\left(\dfrac{(c-a)}{a\cdot c}\cdot x\right)}\ \ \ \ \ \ \textnormal{if}\ \ \ \ a \neq c\\
\nonumber
    -x\ \ \ \ \ \ \ \ \ \ \ \ \ \ \ \ \ \ \ \ \ \ \ \ \ \ \ \ \ \ \ \ \ \ \ \ \ \ \ 
 \textnormal{if}\ \ \ \ a = c
    \end{cases}\\
\\
P_D(x;a, b|c) &=& c \times \begin{cases}
\nonumber
\dfrac{a \cdot b \cdot c}{(a-c)(b-c)}\cdot\exp{\left(\dfrac{(c-b)}{b \cdot c}\cdot x\right)}\ \ \ \ \ \ \textnormal{if}\ \ \ \ c \neq a \ \ \wedge \ \ c\neq b\\
\\
    -\dfrac{a}{(a-c)}\cdot x \ \ \ \ \ \ \ \ \ \ \ \ \ \ \ \ \ \ \ \ \ \ \ \ \ \ \ \ \ \ \ \ \ \ \ \ \ \ \ \ \ 
 \textnormal{if}\ \ \ \ c = b\\
 \\
\nonumber
    \dfrac{b}{(b-c)^2}\cdot \left[(c-b)\cdot x - c\cdot b\right]\cdot \exp{\left( \dfrac{(c-b)}{b\cdot c}\cdot x\right)} \ \ \ \ \ \ \ \ \ \ \ \ \ \ \ \ \ \ \textnormal{if}\ \ \ \ c = a
    \end{cases}\\
\\
S_D(x;a|c) &=& \begin{cases}
 \dfrac{a^2\cdot c^2}{(a-c)^2}\cdot \exp{\left(\dfrac{(c-a)}{a\cdot c}\cdot x\right)}\ \ \ \ \ \ \textnormal{if}\ \ \ \ c \neq a \\
 \\ \nonumber
\dfrac{x^2}{2} \ \ \ \ \ \ \ \ \ \ \ \ \ \ \ \ \ \ \textnormal{if}\ \ \ \ c = a
\end{cases}\\
\end{eqnarray}
Using these definitions the Type Ia SN rates are
\begin{subequations}
\label{Eq_R1a}
\begin{equation}
\begin{split}
    \label{Eq_R1a_g} 
     \mathcal{R}_{Ia}^G(t)= C_{Ia}  A_{G} \sigma' \nu_L\sqrt{\frac{\pi}{2}} \left\lbrace\sum_{\substack{j=1\\ 
                  \tau_j \neq \alpha^{-1}}}^N \dfrac{A_j}{\beta_j} \cdot \exp\left(-\frac{1}{\tau_j}\left[\Delta t_j-\tau'-\frac{\sigma'^2}{2\tau_j}\right]\right)  \cdot \left[\textnormal{erf}\left(\frac{\Lambda_G(\Delta t_j)-\eta_j}{\sqrt{2}\sigma'}\right)-\textnormal{erf}\left(\frac{\lambda_G(\Delta t_j)-\eta_j}{\sqrt{2}\sigma'}\right)\right] \right.
    \\
    \left.
    -\sum_{\substack{j=1\\ 
                  \tau_j \neq \alpha^{-1}}}^N \dfrac{A_j}{\beta_j} \cdot \exp\left(-\alpha\left[\Delta t_j-\tau'-\frac{\sigma'^2\alpha}{2}\right]\right) \cdot \left[\textnormal{erf}\left(\frac{\Lambda_G(\Delta t_j)-\eta_\alpha}{\sqrt{2}\sigma'}\right)-\textnormal{erf}\left(\frac{\lambda_G(\Delta t_j)-\eta_\alpha}{\sqrt{2}\sigma'}\right)\right]\right.
    \\
    \left.
    + \sigma_{gas}(0) \cdot \exp\left(-\alpha\left[t-\tau'-\frac{\sigma'^2\alpha}{2}\right]\right) \cdot \left[\textnormal{erf}\left(\frac{\Lambda_G(t)-\eta_\alpha}{\sqrt{2}\sigma'}\right)-\textnormal{erf}\left(\frac{\lambda_G(t)-\eta_\alpha}{\sqrt{2}\sigma'}\right)\right]\right\rbrace\\
\end{split}
\end{equation}
\begin{equation}
\begin{split}
    \label{Eq_R1a_e}
    \mathcal{R}_{Ia}^E(t) =  C_{Ia} A_{E} \nu_L\left\lbrace \sum_{\substack{j=1\\ 
                  \tau_j \neq \alpha^{-1}}}^N \dfrac{A_j}{\beta_j} \exp\left(-\frac{\Delta t_j}{\tau_j}\right) \left[Q_D\left(\lambda_E(\Delta t_j); \tau_j| \tau_D\right)-Q_D\left(\Lambda_E(\Delta t_j); \tau_j\right| \tau_D)\right]\right.\\
    \left.- \sum_{\substack{j=1\\ 
                  \tau_j \neq \alpha^{-1}}}^N \dfrac{A_j}{\beta_j} \cdot \exp\left(-\alpha \Delta t_j\right) \left[Q_D\left(\lambda_E(\Delta t_j); \alpha^{-1}| \tau_D\right)-Q_D\left(\Lambda_E(\Delta t_j); \alpha^{-1}| \tau_D\right)\right] \right.\\
    \left.+ \sigma_{gas}(0)\cdot \exp\left(-\alpha  t \right) \left[Q_D\left(\lambda_E(t); \alpha^{-1}| \tau_D\right)-Q_D\left(\Lambda_E(t); \alpha^{-1}| \tau_D\right)\right]
    \right.\Bigg\rbrace\\
\end{split}
\end{equation}
\begin{equation}
\begin{split}
    \label{Eq_R1a_i}
     \mathcal{R}_{Ia}^I(t) &=  C_{Ia} A_{I} \tau_I  \nu_L
 \sum_{\substack{j=1\\ 
                  \tau_j \neq \alpha^{-1}}}^N \dfrac{A_j}{\beta_j} \left\lbrace  \exp{\left(-\frac{\Delta t_j-\tau_0}{\tau_j}\right)}  \cdot \left[\textnormal{Ei}\left(\frac{\Lambda_I(\Delta t_j)-\tau_0}{\tau_j}\right)-\textnormal{Ei}\left(\frac{\lambda_I(\Delta t_j)-\tau_0}{\tau_j}\right) \right]+ \right. \\
&+\left. \exp{\left(-\alpha\left[\Delta t_j-\tau_0\right]\right)} \cdot \left[ \textnormal{Ei}\left(\alpha\left[\lambda_I(\Delta t_j)-\tau_0\right]\right) -\textnormal{Ei}\left(\alpha\left[\Lambda_I(\Delta t_j)-\tau_0\right]\right) \right] \right.\Bigg\rbrace \\
&+ C_{Ia} A_{I} \tau_I \nu_L \sigma_{gas}(0) \exp{\left(-\alpha\left[ t-\tau_0\right]\right)} \cdot \left.\bigg[ \ \textnormal{Ei}\left(\alpha\left[\Lambda_I(t)-\tau_0\right]\right)-\textnormal{Ei}\left(\alpha\left[\lambda_I(t)-\tau_0\right]\right)\right.\bigg]
\end{split}
\end{equation}
\end{subequations}
so that the global Type Ia SN rate is $\mathcal{R}_{Ia} = \mathcal{R}_{Ia}^G + \mathcal{R}_{Ia}^E + \mathcal{R}_{Ia}^I$, where $\Delta t_j \equiv t-t_j$. The $\textnormal{Ei}$ function presented in \ref{Eq_R1a_i} refers to the so-called exponential integral, though for computational purposes it is better to use a modified version $\widetilde{\textnormal{Ei}}$ without divergences at the origin:
\begin{equation}
\label{Eq_Ei}
    \widetilde{\textnormal{Ei}}(x)= \begin{cases}
 Ei(x)\equiv \int_{-\infty}^x y^{-1}e^{y}\ dy\ \ \ \ \ \ \textnormal{if}\ \ \ \ x \neq 0 \\
 \\
0 \ \ \ \ \ \ \ \ \ \ \ \ \ \ \ \ \ \ \textnormal{if}\ \ \ \ x = 0
\end{cases}\\
\end{equation}
\par We separate the individual contribution of each term in Eq. \ref{eq:diff_equation2} to the global solution $\sigma_{X}$ as $\sigma_{X}=\sigma_{X,IRA}+\sigma_{X,Ia}$. The term related to the IRA approximation reads
\begin{equation}
\begin{split}
    \label{Eq_sigmaIRA}
    \sigma_{X, IRA}(t)&=\sigma_X(0)\exp{\left(-\alpha t\right)}  + \langle y_X \rangle(1-R)\nu_L\sum_{\substack{j=1\\ 
                  \tau_j \neq \alpha^{-1}}}^N \dfrac{A_j}{\beta_j}  \exp\left({-\alpha \Delta t_j}\right)\theta(\Delta t_j) \left[ \dfrac{\exp\left({\beta_j \Delta t_j}\right)-1 }{\beta_j}-
    {\Delta t_j} \right] + \\ +&\langle y_X \rangle(1-R)\nu_L\sigma_{gas}(0) \cdot t\cdot\exp\left({-\alpha t}\right) \theta(t)
   \end{split}
\end{equation}
\par We can write $\sigma_{X,Ia}$ as the sum of the three components of the DTD as $\sigma_{X, Ia}=\sigma_{X, Ia, G}+\sigma_{X, Ia, E}+\sigma_{X, Ia, I}$. For sake of illustration, we consider the case in which $\sigma_{X, Ia, G}$, $\sigma_{X, Ia, E}$ and $\sigma_{X, Ia, I}$ are defined by a unique Gaussian, exponential and $(t-\tau_0)^{-1}$ distributions, respectively. This allows us to obviate the $i$-index in the parameters of the DTD (see Eq. \ref{Eq_DTD_component}) and simplify the notation. For more realistic DTDs, like the ones considered in this work, it is necessary to compute the associated $\sigma_{X, Ia, G}$, $\sigma_{X, Ia, E}$ or $\sigma_{X, Ia, I}$ of each individual component of the DTD. Thus, the terms $\sigma_{X, Ia, G}$, $\sigma_{X, Ia, E}$ and $\sigma_{X, Ia, I}$ read
\begin{subequations}
\label{Eq_GenSol}
\begin{equation}
\label{Eq_GenSol_DTD_Gauss}
\begin{split}
    \sigma_{X, Ia, G}(t)&=\\
    =&\langle m_{X, Ia} \rangle C_{Ia}A_G\sigma'\nu_L\sqrt{\frac{\pi}{2}}\sum_{\substack{j=1\\ 
                  \tau_j \neq \alpha^{-1}}}^N {\dfrac{A_j}{\beta_j^2}} \theta(\Delta t_j -\tau_1) \exp{\left(-\alpha \Delta t_j + \frac{\tau'}{\tau_j}+\frac{\sigma'^2}{2\tau_j^2}\right)} \cdot \left\lbrace \left[\textnormal{erf}\left(\frac{\tau_1-\eta_{\alpha}}{\sqrt{2}\sigma'}\right)-\textnormal{erf}\left(\frac{ \Lambda_G(\Delta t_j)-\eta_{\alpha}}{\sqrt{2}\sigma'}\right)\right]
    \cdot \right.\\
    &\left.\cdot
    \exp{\left(\beta_j \eta_j+\frac{1}{2}\beta_j^2\sigma'^2\right)}+\left[\textnormal{erf}\left(\frac{\Lambda_G(\Delta t_j)-\eta_j}{\sqrt{2}\sigma'}\right)-\textnormal{erf}\left(\frac{\tau_1-\eta_j}{\sqrt{2}\sigma'}\right)\right]\exp{\left(\beta_j \Lambda_G(\Delta t_j)\right)} 
        \right.\\
    &\left.+
    \theta(\Delta t_j -\tau_2)\left[\exp{\left(\beta_j \Delta t_j\right)}-\exp{\left(\beta_j \tau_2\right)}\right] \left[\textnormal{erf}\left(\frac{\tau_2-\eta_j}{\sqrt{2}\sigma'}\right)-\textnormal{erf}\left(\frac{\tau_1-\eta_j}{\sqrt{2}\sigma'}\right)\right]    \right\rbrace \\ 
    -&\langle m_{X, Ia} \rangle C_{Ia}A_G\sigma'\nu_L\sum_{\substack{j=1\\ 
                  \tau_j \neq \alpha^{-1}}}^N {\dfrac{A_j}{\beta_j}} \theta(\Delta t_j -\tau_1)\exp{\left(-\alpha \Delta t_j+ {\tau'\alpha}+\frac{\sigma'^2\alpha^2}{2}\right)}\cdot \left\lbrace \sqrt{\frac{\pi}{2}}\cdot\left[\left(\Lambda_G(\Delta t_j)-\eta_\alpha\right)\cdot\textnormal{erf}\left(\frac{\Lambda_G(\Delta t_j)-\eta_\alpha}{\sqrt{2}\sigma'}\right)
            \right.\right.\\
    &\left.\left.
    -\left(\tau_1-\eta_\alpha\right)\cdot\textnormal{erf}\left(\frac{\tau_1-\eta_\alpha}{\sqrt{2}\sigma'}\right)+\left(\tau_1-\Lambda_G(\Delta t_j)\right)\cdot\textnormal{erf}\left(\frac{\tau_1-\eta_\alpha}{\sqrt{2}\sigma'}\right)\right]
    \right.+\\
    &\left. 
   \sigma' \left[\exp{\left(-\frac{\left(\Lambda(\Delta t_j)-\eta_\alpha\right)^2}{2 \sigma'^2}\right)}-\exp{\left(-\frac{\left(\tau_1-\eta_\alpha\right)^2}{2 \sigma'^2}\right)}\right]+\sqrt{\frac{\pi}{2}}\theta\left(\Delta t_j - \tau_2\right)\cdot\left(\Delta t_j - \tau_2\right)\left[\textnormal{erf}\left(\frac{\tau_2-\eta_\alpha}{\sqrt{2}\sigma'}\right)-\textnormal{erf}\left(\frac{\tau_1-\eta_\alpha}{\sqrt{2}\sigma'}\right)\right]\right\rbrace + \\ 
    +&\langle m_{X, Ia} \rangle C_{Ia}A_G\sigma'\nu_L \sigma_{gas}(0)\cdot \theta( t -\tau_1)\exp{\left(-\alpha t+ {\tau'\alpha}+\frac{\sigma'^2\alpha^2}{2}\right)}\cdot \left\lbrace \sqrt{\frac{\pi}{2}}\cdot\left[\left(\Lambda_G(t)-\eta_\alpha\right)\cdot\textnormal{erf}\left(\frac{\Lambda_G(t)-\eta_\alpha}{\sqrt{2}\sigma'}\right)
            \right.\right.\\
    &\left.\left.
    -\left(\tau_1-\eta_\alpha\right)\cdot\textnormal{erf}\left(\frac{\tau_1-\eta_\alpha}{\sqrt{2}\sigma'}\right)+\left(\tau_1-\Lambda_G(t)\right)\cdot\textnormal{erf}\left(\frac{\tau_1-\eta_\alpha}{\sqrt{2}\sigma'}\right)\right]
    \right.+\sigma' \left[\exp{\left(-\frac{\left(\Lambda_G(t)-\eta_\alpha\right)^2}{2 \sigma'^2}\right)}-\exp{\left(-\frac{\left(\tau_1-\eta_\alpha\right)^2}{2 \sigma'^2}\right)}\right]+\\
    &\left. 
    \theta\left(t - \tau_2\right)\cdot\left(t - \tau_2\right)\sqrt{\frac{\pi}{2}}\cdot\left[\textnormal{erf}\left(\frac{\tau_2-\eta_\alpha}{\sqrt{2}\sigma'}\right)-\textnormal{erf}\left(\frac{\tau_1-\eta_\alpha}{\sqrt{2}\sigma'}\right)\right]\right\rbrace
    \end{split}
\\
\end{equation}
\begin{equation}
\label{Eq_GenSol_DTD_Exp}
\begin{split}
\sigma_{X, Ia, E}(t)&=\\
=&\langle m_{X, Ia} \rangle C_{Ia}A_E \nu_L\sum_{\substack{j=1\\ 
                  \tau_j \neq \alpha^{-1}}}^N \dfrac{A_j}{\beta_j} \exp{\left(-\alpha \Delta t_j\right)}\left\lbrace 
Q_D\left(\tau_2; \tau_j \left| \tau_D \right.\right) \left[Q_D\left(\Delta t_j; \alpha^{-1} \left| \tau_j \right.\right)-Q_D\left(\Lambda_E(\Delta t_j); \alpha^{-1} \left| \tau_j \right.\right)\right] \right.\\
+&\left. Q_D\left(\tau_1; \tau_j \left| \tau_D \right.\right) \left[Q_D\left(\lambda_E(\Delta t_j); \alpha^{-1} \left| \tau_j \right.\right)-Q_D\left(\Delta t_j; \alpha^{-1} \left| \tau_j \right.\right)\right] 
+ Q_D\left(\tau_2; \alpha^{-1} \left| \tau_D \right.\right) \left[Q_D\left(\Lambda_E(\Delta t_j); \alpha^{-1} \left| \alpha^{-1} \right.\right)-Q_D\left(\Delta t_j; \alpha^{-1} \left| \alpha^{-1} \right.\right)\right]  \right.\\
+&\left. Q_D\left(\tau_1; \alpha^{-1} \left| \tau_D \right.\right) \left[Q_D\left(\Delta t_j; \alpha^{-1} \left| \alpha^{-1} \right.\right)-Q_D\left(\lambda_E(\Delta t_j); \alpha^{-1} \left| \alpha^{-1} \right.\right)\right]\right.\\
+&\left. P_D\left(\Lambda_E(\Delta t_j);\tau_j, \alpha^{-1}\left|\tau_D\right.\right)-S_D\left(\Lambda_E(\Delta t_j);\alpha^{-1}\left|\tau_D\right.\right)-P_D\left(\lambda_E(\Delta t_j);\tau_j, \alpha^{-1}\left|\tau_D\right. \right)+S_D\left(\lambda_E(\Delta t_j);\alpha^{-1}\left|\tau_D\right. \right)\right\rbrace\\
+&\langle m_{X, Ia} \rangle C_{Ia}A_E \nu_L \sigma_{gas}(0)\cdot\exp{\left(-\alpha t\right)}\left\lbrace 
Q_D\left(\tau_2; \alpha^{-1} \left| \tau_D \right.\right) \left[Q_D\left(t; \alpha^{-1} \left| \alpha^{-1} \right.\right)-Q_D\left(\Lambda_E(t); \alpha^{-1} \left| \alpha^{-1} \right.\right)\right]  \right.\\
-&\left. Q_D\left(\tau_1; \alpha^{-1} \left| \tau_D \right.\right) \left[Q_D\left(t; \alpha^{-1} \left| \alpha^{-1} \right.\right)-Q_D\left(\lambda_E(t); \alpha^{-1} \left| \alpha^{-1} \right.\right)\right]
+S_D\left(\Lambda_E(t);\alpha^{-1}\left|\tau_D\right.\right)-S_D\left(\lambda_E(t);\alpha^{-1}\left|\tau_D\right. \right)\right\rbrace\\
\end{split}
\end{equation}
\begin{equation}
\label{Eq_GenSol_DTD_Inv}
\begin{split}
\sigma_{X, Ia, I}(t)&=\\
=&\langle m_{X, Ia} \rangle C_{Ia}A_I \tau_I \nu_L\sum_{\substack{j=1\\ 
                  \tau_j \neq \alpha^{-1}}}^N \dfrac{A_j}{\beta_j^2} \exp{\left(-\alpha\left[\Delta t_j-\tau_0\right]\right)}\left\lbrace \theta\left(\Delta t_j-\tau_1\right) \cdot \left[\textnormal{Ei}\left(\frac{\Lambda_I(\Delta t_j)-\tau_0}{\tau_j}\right)\cdot \exp{\left(\beta_j\left[\Lambda_I(\Delta t_j)-\tau_0\right]\right)}+\right.\right.\\
&\left.\left.+\textnormal{Ei}\left(\alpha[\tau_1-\tau_0]\right)-\textnormal{Ei}\left(\alpha\left[\Lambda_I(\Delta t_j)-\tau_0\right]\right)-\textnormal{Ei}\left(\frac{\tau_1-\tau_0}{\tau_j}\right)\cdot\exp{\left(\beta_j\left[\Delta t_j-\tau_0\right]\right)} \right] \right. + \\
&\left.+ \theta\left(\Delta t_j -\tau_2\right)  \cdot \left[\exp{\left(\beta_j \left[\Delta t_j-\tau_0\right]\right)}-\exp{\left(\beta_j \left[\tau_2-\tau_0\right]\right)}\right]\cdot\textnormal{Ei}\left(\dfrac{\tau_2-\tau_0}{\tau_j}\right)\right\rbrace \\
-&\langle m_{X, Ia} \rangle C_{Ia}A_I \tau_I \nu_L\sum_{\substack{j=1\\ 
                  \tau_j \neq \alpha^{-1}}}^N \dfrac{A_j}{\beta_j} \exp{\left(-\alpha\left[\Delta t_j-\tau_0\right]\right)} \left\lbrace \theta\left(\Delta t_j -\tau_1\right)\left[ \left(\Lambda_I(\Delta t_j)-\tau_0\right)\cdot\textnormal{Ei}\left(\alpha\left[\Lambda_I(\Delta t_j)-\tau_0\right]\right) + \right. \right.\\
+&\left.\left.\frac{\exp{\left(\alpha\left[\tau_1-\tau_0\right]\right)}-\exp{\left(\alpha\left[\Lambda_I(\Delta t_j)-\tau_0\right]\right)}}{\alpha} -\left(\Delta t_j-\tau_0\right)\cdot\textnormal{Ei}\left(\alpha[\tau_1-\tau_0]\right) \right] + \theta\left(\Delta t_j -\tau_2\right)\cdot\textnormal{Ei}\left(\alpha\left[\tau_2-\tau_0\right]\right)\cdot\left( \Delta t_j-\tau_2\right) \right.\Bigg\rbrace+\\
+&\langle m_{X, Ia} \rangle C_{Ia}A_I \tau_I \nu_L \sigma_{gas}(0) \exp{\left(-\alpha [t-\tau_0]\right)} \left\lbrace \theta\left(t -\tau_1\right)\left[ \left(\Lambda_I(t)-\tau_0\right)\cdot\textnormal{Ei}\left(\alpha[\Lambda_I(t)-\tau_0]\right) - \frac{\exp{\left(\alpha[\Lambda_I(t)-\tau_0]\right)}}{\alpha} \right. \right. +\\
+&\left. \left. \frac{\exp{\left(\alpha[\tau_1-\tau_0]\right)}}{\alpha} -(t-\tau_0)\cdot\textnormal{Ei}\left(\alpha[\tau_1-\tau_0]\right)\right]+ \theta\left(t -\tau_2\right)\cdot\textnormal{Ei}\left(\alpha\left[\tau_2-\tau_0\right]\right)\cdot\left( t-\tau_2\right) \right.\Bigg\rbrace\\
\end{split}
\end{equation}
\end{subequations}
\subsection{Case $\tau_j=\alpha^{-1}$}\label{app_sol_peculiar}
\par According to Eq. \ref{infall_eq}, an infall with a timescale $\tau_j$ equal to $\alpha^{-1}$ contributes to the SFR with a different functional form than the more general case $\tau_j\neq\alpha^{-1}$. This discrepancy implies the inclusion of additional terms to the Type Ia SN rates $\mathcal{R}_{\text{Ia}}(t)$ (Eqs. \ref{Eq_R1a}) and to the solutions $\sigma_{X,IRA}$ and $\sigma_{X,Ia}$ (\ref{Eq_sigmaIRA} and \ref{Eq_GenSol}, respectively). Though not used in this work, we include these extra terms for completeness. Defining the functions $\widetilde{Q}_D$ and $\widetilde{S}_D$ as
\begin{eqnarray}
\label{Functions}
    \widetilde{Q}_D(x,y;a|c) &=& \begin{cases}
\dfrac{a \cdot c}{(a-c)^2}\cdot{\left[\exp{\left(\dfrac{c-a}{c\cdot a}x\right)}\left[c\cdot a + (c-a)\cdot(y-x)\right]\right]}\ \ \ \ \ \ \textnormal{if}\ \ \ \ a \neq c\\
\nonumber \\
\nonumber
    -\dfrac{\left(y-x\right)^2}{2}\ \ \ \ \ \ \ \ \ \ \ \ \ \ \ \ \ \ \ \ \ \ \ \ \ \ \ \ \ \ \ \ \ \ \ \ \ \ \ 
 \textnormal{if}\ \ \ \ a = c
    \end{cases}\\
\\
\widetilde{S}_D(x;a|c) &=& \begin{cases}
 \dfrac{(c\cdot a)^3}{(c-a)^3}\cdot\left[ -1 + \exp{\left(\dfrac{(a-c)}{a\cdot c}\cdot x\right)}\left(1+\dfrac{(c-a)}{c\cdot a}\cdot x + \dfrac{(c-a)^2}{2(c\cdot a)^2}\cdot x^2\right)\right]\theta(x)\ \ \ \ \ \ \textnormal{if}\ \ \ \ c \neq a \\
 \\ \nonumber
-\dfrac{x^3}{6}\theta(x) \ \ \ \ \ \ \ \ \ \ \ \ \ \ \ \ \ \ \textnormal{if}\ \ \ \ c = a
\end{cases}\\
\end{eqnarray}
the corrections $\Delta \mathcal{R}_{\text{Ia}}$ to the Type Ia SN rates read
\begin{subequations}
\label{Eq_R1a_peculiar}
\begin{equation}
\begin{split}
    \label{Eq_R1a_g_peculiar}
     \Delta \mathcal{R}_{Ia}^G(t)&= C_{Ia}  A_{G} \sigma' \nu_L \sum_{\substack{j=1\\ 
                  \tau_j = \alpha^{-1}}}^N A_j \cdot \exp{\left(-\alpha\left[\Delta t_j-\tau'-\frac{\alpha \sigma'^2}{2}\right]\right)}  \cdot \left\lbrace \sqrt{\frac{\pi}{2}}\cdot\left(\Delta t_j-\eta_{\alpha}\right)\cdot\left[\textnormal{erf}\left(\frac{\Lambda_G(\Delta t_j)-\eta_{\alpha}}{\sqrt{2}\sigma'}\right)-\textnormal{erf}\left(\frac{\lambda_G(\Delta t_j)-\eta_{\alpha}}{\sqrt{2}\sigma'}\right)\right] + \right. \\
                  +& \left. \sigma'\left[\exp{\left(-\dfrac{\left[\Lambda_G(\Delta t_j)-\eta_{\alpha}\right]^2}{2\sigma'^2}\right)}-\exp{\left(-\dfrac{\left[\lambda_G(\Delta t_j)-\eta_{\alpha}\right]^2}{2\sigma'^2}\right)}\right]\right\rbrace
\end{split}
\end{equation}
\begin{equation}
\begin{split}
    \label{Eq_R1a_e_peculiar}
    \Delta\mathcal{R}_{Ia}^E(t) =  C_{Ia} A_{E} \nu_L \sum_{\substack{j=1\\ 
                  \tau_j = \alpha^{-1}}}^N A_j \exp\left(-\alpha \Delta t_j\right) \cdot \left[\widetilde{Q}_D\left(\Lambda_E(\Delta t_j), \Delta t_j; \alpha^{-1}| \tau_D\right)-\widetilde{Q}_D\left(\lambda_E(\Delta t_j), \Delta t_j; \alpha^{-1}| \tau_D\right)\right]
\end{split}
\end{equation}
\begin{equation}
\begin{split}
    \label{Eq_R1a_i_peculiar}
     \Delta \mathcal{R}_{Ia}^I(t) &=  C_{Ia} A_{I} \tau_I \nu_L 
 \sum_{\substack{j=1\\ 
                  \tau_j = \alpha^{-1}}}^N A_j \Bigg\lbrace  \left(\Delta t_j -\tau_0\right)\cdot \exp{\left(-\alpha\left[\Delta t_j-\tau_0\right]\right)}  \cdot \left[\textnormal{Ei}\left(\alpha\left[\Lambda_I(\Delta t_j)-\tau_0\right]\right)-\textnormal{Ei}\left(\alpha\left[\lambda_I(\Delta t_j)-\tau_0\right] \right) \right] + \\
+& \left. \dfrac{ \exp{\left(\alpha[\lambda_I (\Delta t_j)-\Delta t_j]\right)}- \exp{\left(\alpha[\Lambda_I (\Delta t_j)-\Delta t_j]\right)}}{\alpha} \right\rbrace
\end{split}
\end{equation}
\end{subequations}
\par Similarly, the IRA solution requires the addition of the term
%
%
\begin{equation}
    \label{Eq_sigmaIRA_peculiar}
    \Delta \sigma_{X, IRA}(t)= \langle y_X \rangle(1-R)\cdot\frac{\nu_L}{2}\sum_{\substack{j=1\\ 
                  \tau_j = \alpha^{-1}}}^N A_j (\Delta t_j)^2  \exp\left({-\alpha \Delta t_j}\right)\theta(\Delta t_j)
\end{equation}
\par while for the Type Ia SN enrichment the contribution of each DTD is 
\begin{subequations}
\label{Eq_GenSol_peculiar}
\begin{equation}
\begin{split}
    \label{Eq_GenSol_DTD_Gauss_peculiar}
     \Delta \sigma_{X,Ia,G}(t)&= \langle m_{X, Ia} \rangle C_{Ia}  A_{G} \sigma' \nu_L \sqrt{\frac{\pi}{2}} \cdot \sum_{\substack{j=1\\ 
                  \tau_j = \alpha^{-1}}}^N A_j \cdot \exp\left(-\alpha\left[\Delta t_j-\tau'-\dfrac{\alpha\sigma'^2}{2}\right]\right)  \cdot \left\lbrace \left(\Delta t_j - \eta_{\alpha}\right)\cdot\left[\textnormal{erf}\left(\frac{\Lambda_G(\Delta t_j)-\eta_{\alpha}}{\sqrt{2}\sigma'}\right)-\right.\right. \\
                   & \left.\left.
                  \textnormal{erf}\left(\frac{\lambda_G(\Delta t_j)-\eta_{\alpha}}{\sqrt{2}\sigma'}\right)\right] + \sigma'\cdot\left[\exp{\left(-\dfrac{\left(\Lambda_G(\Delta t_j)-\eta_{\alpha}\right)^2}{2\sigma'^2}\right)}-\exp{\left(-\dfrac{\left(\lambda_G(\Delta t_j)-\eta_{\alpha}\right)^2}{2\sigma'^2}\right)}\right]\right\rbrace
\end{split}
\end{equation}
\begin{equation}
\begin{split}
    \label{Eq_GenSol_DTD_Expo_peculiar}
     \Delta \sigma_{X,Ia,E}(t)&= \langle m_{X, Ia} \rangle C_{Ia}  A_{E} \nu_L \sum_{\substack{j=1\\ 
                  \tau_j = \alpha^{-1}}}^N A_j \cdot \exp{\left(-\dfrac{\Delta t_j}{\tau_D}\right)}\cdot  \left[\widetilde{S}_D\left(\Delta t_j-\Lambda_E(\Delta t_j); \alpha^{-1}|\tau_D\right)-\widetilde{S}_D\left(\Delta t_j-\lambda_E(\Delta t_j); \alpha^{-1}|\tau_D\right)\right]
\end{split}
\end{equation}
\begin{equation}
\begin{split}
    \label{Eq_GenSol_DTD_Inv_peculiar}
     \Delta \sigma_{X,Ia,I}(t)&= \langle m_{X, Ia} \rangle C_{Ia}  A_{I}\tau_{I} \nu_L \sum_{\substack{j=1\\ 
                  \tau_j = \alpha^{-1}}}^N A_j \cdot \exp{\left(-\alpha\left[\Delta t_j-\tau_0\right]\right)}\cdot \theta\left(\Delta t_j-\tau_1 \right)\cdot \left\lbrace 
\dfrac{\left(\Lambda_I(\Delta t_j)-\tau_0\right)^2}{2} \textnormal{Ei}\left(\alpha\left[\Lambda_I(\Delta t_j)-\tau_0\right]\right)-\right. \\
& \left.
\dfrac{\left(\Delta t_j-\tau_0\right)^2}{2} \textnormal{Ei}\left(\alpha\left[\tau_1-\tau_0\right]\right)+
\dfrac{\exp{\left(\alpha\left[\tau_1-\tau_0\right]\right)}-\exp{\left(\alpha\left[\Lambda_I(\Delta t_j)-\tau_0\right]\right)}}{2\alpha^2}+\right. \\
+& \left.
\dfrac{1}{2\alpha}\left[\left(\tau_1-\tau_0\right)\exp{\left(\alpha\left[\tau_1-\tau_0\right]\right)}-\left(\Lambda_I(\Delta t_j)-\tau_0\right)\exp{\left(\alpha\left[\Lambda_I(\Delta t_j)-\tau_0\right]\right)}\right]+
\dfrac{1}{\alpha}\left(\Delta t_j-\tau_1\right)\exp{\left(\alpha\left[\tau_1-\tau_0\right]\right)}\right. \\
+& \left.
\theta\left(\Delta t_j-\tau_2\right)\left[\dfrac{\textnormal{Ei}\left(\alpha\left[\tau_2-\tau_0\right]\right)}{2}\cdot\left(\left[\Delta t_j-\tau_0\right]^2-\left[\tau_2-\tau_0\right]^2\right)-\dfrac{\left(\Delta t_j-\tau_2\right)}{\alpha}\exp{\left(\alpha\left[\tau_2-\tau_0\right]\right)}\right]
\right\rbrace
\end{split}
\end{equation}
\end{subequations}
\newpage
\section{Two infall model with the CLOSE G05 DTD}\label{Sec_CLOSEG05}
\par
\begin{figure*}[!h]
    \centering
    \includegraphics[width=0.99\textwidth]{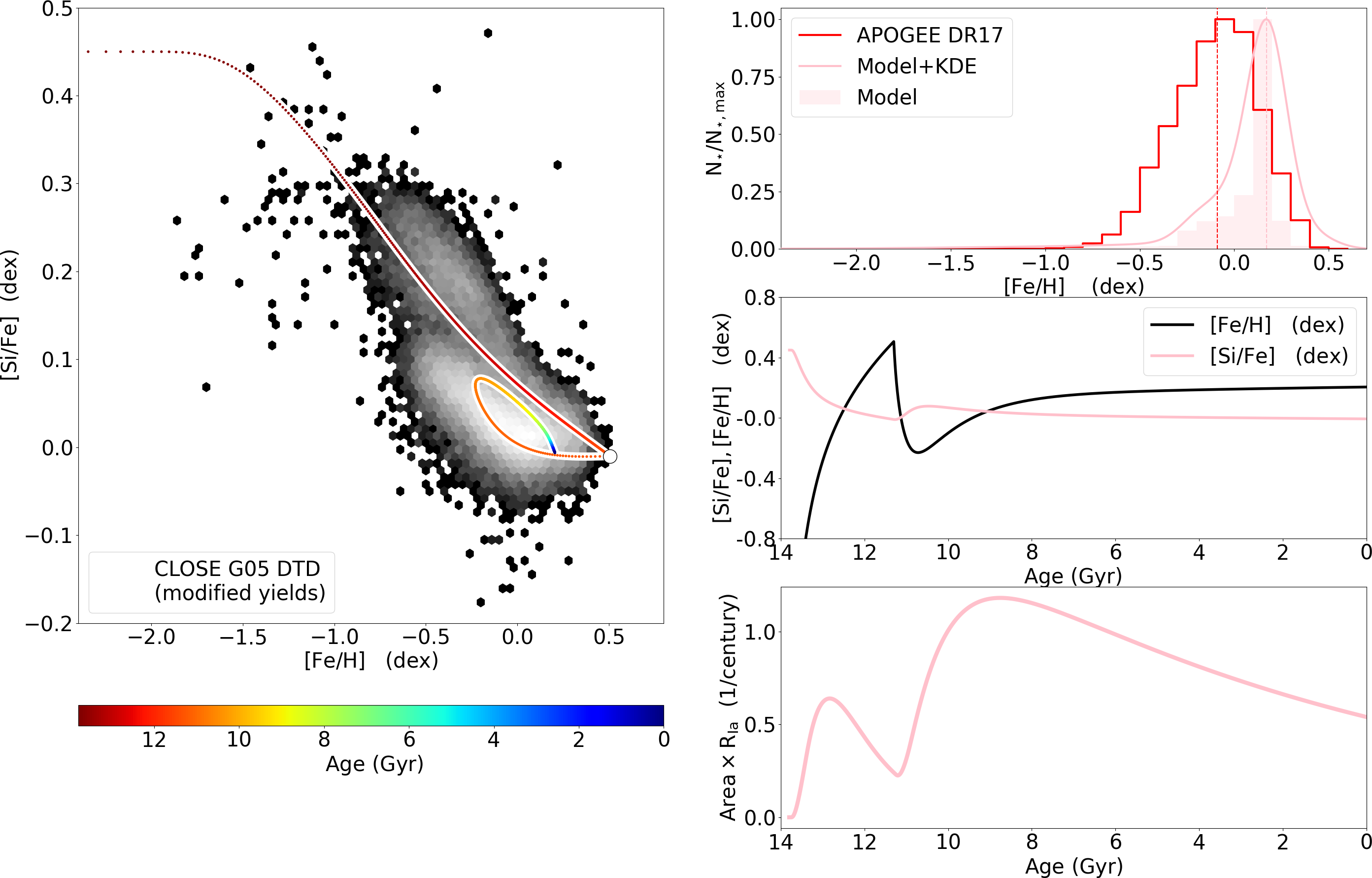}
    \caption{Similar to Figs. \ref{Fig_ChemSpace_2infall} and \ref{Fig_ChemSpace_2infall_specialyields_WideG05} but assuming the CLOSE G05 DTD. The yields $\langle y_{\textnormal{Si}}\rangle$ and $\langle m_{\textnormal{Si},Ia}\rangle$ for silicon have been re-scaled by a factor of 95\%. The infall and model parameters are: $\tau_1=0.5$~Gyr, $\tau_2=8.75$~Gyr, $t_1=0$~Gyr, $t_2=2.5$~Gyr, $A_1\approx10.227$~$\rm M_{\odot}\,pc^{-2}\, Gyr^{-1}$ and $A_2\approx6.004$~$\rm M_{\odot}\,pc^{-2}\, Gyr^{-1}$ (equivalent to a mass ratio between infalls of $7.5$), $\omega=0.2$ and $\nu_L=1.5$~$Gyr^{-1}$.}
    \label{Fig_ChemSpace_2infall_specialyields_CloseG05}
\end{figure*}

 In this section, we illustrate in Fig. \ref{Fig_ChemSpace_2infall_specialyields_CloseG05} the chemical evolution of silicon and iron for the two infall model assuming the CLOSE G05 DTD.
\end{appendix}
\end{document}